\documentclass[aps,prx,10pt,twocolumn,showpacs,preprintnumbers,amsmath,amssymb,superscriptaddress]{revtex4-2}%
\usepackage{lineno}
\usepackage{graphicx}  
\usepackage{dcolumn}   
\usepackage{bm}        
\usepackage{amssymb}   
\usepackage{amsmath}
\usepackage{blkarray, multirow, graphicx, diagbox, color, xcolor, colortbl}
\usepackage[caption=false]{subfig}
\usepackage{bbm, bbold}
\usepackage{ifthen}
\usepackage[colorlinks, linkcolor = blue, citecolor = blue, filecolor = black, urlcolor = blue]{hyperref}
\usepackage{xkeyval}
\usepackage{moreverb}
\usepackage{rotating}
\usepackage{slashbox}
\usepackage{xspace}
\usepackage{nicefrac}
\usepackage[]{units}
\usepackage{physics}
\usepackage{braket}
\usepackage[inline]{enumitem}
\usepackage{tabto}
\usepackage{listings}
\usepackage{xstring}
\usepackage{glossaries}
\usepackage{etoolbox}
\usepackage[makeroom]{cancel}
\usepackage{hyphenat}
\usepackage{booktabs}

\glsdisablehyper

\def\ReplaceStr#1{%
	\IfSubStr{#1}{p}{%
		\StrSubstitute{#1}{p}{.}}{#1}}

\newcommand\subfigref[1]{\protect\subref{#1}}

\usepackage[customcolors]{hf-tikz} 
\usepackage{tikz}
\usepackage{calc}
\usetikzlibrary{external}
\graphicspath{{figures/autogen/}}
\usepackage{pgffor}
\usepackage{pgfplots}
\pgfplotsset{compat=newest}
\usepackage{pgfplotstable}
\usepgfplotslibrary{groupplots}
\usepgfplotslibrary{fillbetween}

\tikzstyle{n} = [draw,shape=ellipse,minimum size=1.5em,inner sep=0pt,fill=white!20, minimum width=2.5em]
\tikzstyle{Init} = [n,color=green,fill=green!20,text=black]
\tikzstyle{Fin} = [n,color=red,fill=red!20,text=black]
\tikzstyle{Ghost} = [minimum size=1.5em,inner sep=0pt,color=white,text=black]
\tikzstyle{Multiple} = [draw,shape=rect,minimum size=2em,inner sep=0pt]

\tikzstyle{ghostA} = [text=red!70,thick, minimum size=2*(5pt-\pgflinewidth), inner sep=0pt, outer sep=0pt]
\tikzstyle{ghostB} = [text=blue!70,thick, minimum size=2*(5pt-\pgflinewidth), inner sep=0pt, outer sep=0pt]
\tikzstyle{siteA} = [draw=red!70,circle,thick, minimum size=2*(5pt-\pgflinewidth), inner sep=0pt, outer sep=0pt]
\tikzstyle{siteB} = [draw=blue!70,circle,thick, minimum size=2*(5pt-\pgflinewidth), inner sep=0pt, outer sep=0pt]
\tikzstyle{operatorA} = [cross out, draw=red!70, thick, minimum size=2*(5pt-\pgflinewidth), inner sep=0pt, outer sep=0pt]
\tikzstyle{operatorB} = [cross out, draw=blue!70, thick, minimum size=2*(5pt-\pgflinewidth), inner sep=0pt, outer sep=0pt]

\tikzstyle{site} = [circle,thick,inner sep=0.2pt,minimum width=1.25em,font=\footnotesize,draw=blue!50!white,fill=blue!15!white,text opacity=1]
\tikzstyle{unsite} = [circle, outer sep=0pt,inner sep=0.2pt,minimum width=1.25em]
\tikzstyle{ghost} = []
\tikzstyle{op} = [regular polygon, regular polygon sides=4, draw=orange!50, fill=orange!20, thick, inner sep=0.2pt, minimum width=1.25em, minimum height=1.5em,font=\footnotesize]
\tikzstyle{ld} = [inner sep=1pt, font=\small]
\tikzstyle{intersite} = [regular polygon, regular polygon sides=4, shape border rotate= 45, draw=black!50,fill=black!20,thick,inner sep=0pt,minimum width=1.5em]
\usetikzlibrary{decorations.pathreplacing, calligraphy}

\definecolor{colorA}{rgb} {0.48,0,0.5275}
\definecolor{colorB}{rgb} {0.11,0.663,0.51}
\definecolor{colorC}{rgb} {0.3373,0.7059,0.9137}
\definecolor{colorD}{rgb} {0.902,0.8735,0.1}
\definecolor{colorE}{rgb} {0.9451,0.902,0.3255}
\definecolor{colorF}{rgb} {0.3373,0.3255,0.902}
\definecolor{colorG}{rgb} {0.9451,0.3255,0.3373}
\definecolor{colorH}{rgb} {0.11,0.3255,0.3373}

\usetikzlibrary
{
	calc,
	decorations,
	pgfplots.patchplots,
	plotmarks,
	patterns,
	positioning,
	petri,
	arrows,
	intersections,
	decorations.markings,
	backgrounds,
	fit,
	matrix,
	graphs,
	shapes.geometric,
	decorations.pathreplacing, 
	decorations.pathmorphing,
	shapes.misc,
	shapes.multipart,
	shapes,
	through,
	tikzmark,
	fadings,
}

\pgfplotsset{
	cycle from colormap manual style/.style={
		x=3cm,y=10pt,ytick=\empty,
		stack plots=y,
		every axis plot/.style={line width=2pt},
	},
}

\tikzset{->-/.style={decoration={
			markings,
			mark=at position .5 with {\arrow{>}}},postaction={decorate}}}

\tikzset{-<-/.style={decoration={
			markings,
			mark=at position .5 with {\arrow{>}}},postaction={decorate}}}

\tikzstyle{orientedsnake} = [
decorate, 
decoration={snake},
->
]  
\tikzstyle{orientedshortarrow} = [
decoration={markings,
	mark=at position .33 with {\arrow{>}}},
postaction={decorate}
]  
\tikzstyle{orientedlongarrow} = [
decoration={markings,
	mark=at position .67 with {\arrow{>}}},
postaction={decorate}
]
\tikzset{dbl/.style={double,
		double equal sign distance,
		-implies,
		shorten >=10pt,
		shorten <=10pt}}
\tikzset{
	between/.style args={#1 and #2}{
		at = ($(#1)!0.5!(#2)$)
	}
}
\pgfmathdeclarefunction{linearFct}{2}%
{%
	\pgfmathparse{#1*x+#2}%
}
\pgfmathdeclarefunction{logFct}{2}%
{%
	\pgfmathparse{#1*log10(x)+#2}%
}
\pgfmathdeclarefunction{algebraicFct}{4}%
{%
	\pgfmathparse{#1*x^(#2)+#3/x+#4}%
}
\pgfmathdeclarefunction{specialFct}{4}%
{%
	\pgfmathparse{#1*x^(#2)*exp(#3/x)+#4}%
}
\pgfmathdeclarefunction{pointmetalog}{3}%
{%
	#1
}%

\newcommand{\nodagger}[0]{{\phantom{\dagger}}}

\newcommand{\pprime}[0]{{\prime\prime}}
\newcommand{\overbar}[1]{\mkern 1.5mu\overline{\mkern-1.5mu#1\mkern-1.5mu}\mkern 1.5mu}

\pgfmathdeclarefunction{peierlspotential}{6}%
{
	\pgfmathparse
	{
		#2 / (#3 * #5) 
		* sin(deg(#1 * #3 - #3 * #5 * (x)))
		* exp(-( ( #1 - #5 * ((x) - #4) ) )^2.0 / ( #6^2.0 ) )
	}%
}

\pgfmathdeclarefunction{theoreticalLimit}{2}
{
	\pgfmathparse
	{
		#2*(#1-1.0)*(#1*1.0-(#2*(#1-1.0)+1.0))/(#1*1.0)+0*x
	}
}

\newboolean{buildtikzpics}
\setboolean{buildtikzpics}{false}

\usepackage{ulem}
\usepackage{cleveref}
\Crefname{appendix}{Appendix}{Appendices}
\Crefname{equation}{Equation}{Equations}
\Crefname{figure}{Figure}{Figures}
\Crefname{section}{Section}{Sections}
\Crefname{tabular}{Tabular}{Tabulars}
\crefname{appendix}{App.}{Apps.}
\crefname{equation}{Eq.}{Eqs.}
\crefname{figure}{Fig.}{Figs.}
\crefname{section}{Sec.}{Secs.}
\crefname{tabular}{Tab.}{Tabs.}

\tikzset{>=stealth}

\pgfplotsset{
	compat=1.12,
	/pgf/declare function={
		cos2(\x) = cos(deg(\x*pi));
		g_plus(\x,\t,\s) = (\s-sqrt(\s^2+\t^2*cos2(\x)^2))/(\t*cos2(\x));
		g_minus(\x,\t,\s) = (\s+sqrt(\s^2+\t^2*cos2(\x)^2))/(\t*cos2(\x));
		gp(\x,\t) = ((\x - sqrt(\x^2 + \t^2))/\t); 
		gm(\x,\t) = ((\x + sqrt(\x^2 + \t^2))/\t);
	} 
}

\pgfplotsset{%
	colormap={parula}{%
		rgb=(0.2081,0.1663,0.5292)rgb=(0.2116,0.1898,0.5777)rgb=(0.2123,0.2138,0.627)
		rgb=(0.2081,0.2386,0.6771)rgb=(0.1959,0.2645,0.7279)rgb=(0.1707,0.2919,0.7792)
		rgb=(0.1253,0.3242,0.8303)rgb=(0.0591,0.3598,0.8683)rgb=(0.0117,0.3875,0.882)
		rgb=(0.006,0.4086,0.8828) rgb=(0.0165,0.4266,0.8786)rgb=(0.0329,0.443,0.872)
		rgb=(0.0498,0.4586,0.8641)rgb=(0.0629,0.4737,0.8554)rgb=(0.0723,0.4887,0.8467)
		rgb=(0.0779,0.504,0.8384) rgb=(0.0793,0.52,0.8312)  rgb=(0.0749,0.5375,0.8263)
		rgb=(0.0641,0.557,0.824)  rgb=(0.0488,0.5772,0.8228)rgb=(0.0343,0.5966,0.8199)
		rgb=(0.0265,0.6137,0.8135)rgb=(0.0239,0.6287,0.8038)rgb=(0.0231,0.6418,0.7913)
		rgb=(0.0228,0.6535,0.7768)rgb=(0.0267,0.6642,0.7607)rgb=(0.0384,0.6743,0.7436)
		rgb=(0.059,0.6838,0.7254) rgb=(0.0843,0.6928,0.7062)rgb=(0.1133,0.7015,0.6859)
		rgb=(0.1453,0.7098,0.6646)rgb=(0.1801,0.7177,0.6424)rgb=(0.2178,0.725,0.6193)
		rgb=(0.2586,0.7317,0.5954)rgb=(0.3022,0.7376,0.5712)rgb=(0.3482,0.7424,0.5473)
		rgb=(0.3953,0.7459,0.5244)rgb=(0.442,0.7481,0.5033) rgb=(0.4871,0.7491,0.484)
		rgb=(0.53,0.7491,0.4661)  rgb=(0.5709,0.7485,0.4494)rgb=(0.6099,0.7473,0.4337)
		rgb=(0.6473,0.7456,0.4188)rgb=(0.6834,0.7435,0.4044)rgb=(0.7184,0.7411,0.3905)
		rgb=(0.7525,0.7384,0.3768)rgb=(0.7858,0.7356,0.3633)rgb=(0.8185,0.7327,0.3498)
		rgb=(0.8507,0.7299,0.336) rgb=(0.8824,0.7274,0.3217)rgb=(0.9139,0.7258,0.3063)
		rgb=(0.945,0.7261,0.2886) rgb=(0.9739,0.7314,0.2666)rgb=(0.9938,0.7455,0.2403)
		rgb=(0.999,0.7653,0.2164) rgb=(0.9955,0.7861,0.1967)rgb=(0.988,0.8066,0.1794)
		rgb=(0.9789,0.8271,0.1633)rgb=(0.9697,0.8481,0.1475)rgb=(0.9626,0.8705,0.1309)
		rgb=(0.9589,0.8949,0.1132)rgb=(0.9598,0.9218,0.0948)rgb=(0.9661,0.9514,0.0755)
		rgb=(0.9763,0.9831,0.0538)
	}
}

\newcommand{\splittedTableHeader}[2]%
{%
	\tikzset{external/export next=false}%
	\begin{tikzpicture}%
		\node[anchor=south west, inner sep = 0, outer sep = 0] (n) at (0,0) {\tiny #1};%
		\node[anchor=north east, inner sep = 0, outer sep = 0] (d) at (0,0) {\tiny #2};%
		\node[fit = (n) (d), inner sep = 0, outer sep = 0] (frame) {};%
		\draw[-] (frame.north west) -- (frame.south east);%
	\end{tikzpicture}%
}%

\newcommand{\centeredSplittedTableHeader}[2]%
{%
	\noindent\parbox[c]{\widthof{\splittedTableHeader{#1}{#2}}}%
	{\splittedTableHeader{#1}{#2}}%
}%
\newacronym[shortplural={MPS}]{MPS}{MPS}{matrix\hyp product state}
\newacronym{MPO}{MPO}{matrix-product operator}
\newacronym{SVD}{SVD}{singular-value decomposition}
\newacronym{QCS}{QCS}{quantum-computer simulator}
\newacronym{FSM}{FSM}{finite-state machine}
\newacronym{ACA}{ACA}{adaptive cross approximation}
\newacronym{1D}{1D}{one\hyp dimensional}
\newacronym{QC}{QC}{quantum computer}
\newacronym{CDW}{CDW}{charge\hyp density wave}
\newacronym{SDW}{SDW}{spin\hyp density wave}
\newacronym{DLCZ}{DLCZ}{Duan\hyp Lukin\hyp Cirac\hyp Zoller}
\newacronym{ARPES}{ARPES}{angle-resolved photoemission spectroscopy}
\newacronym{OBC}{OBC}{open-boundary conditions}
\newacronym{PBC}{PBC}{periodic-boundary conditions}
\newacronym{TEBD}{TEBD}{time-evolution block-decimation}
\newacronym{TDVP}{TDVP}{time\hyp dependent variational principle}
\newacronym{iff}{iff}{if and only if}
\newacronym{TC}{TC}{Tavis-Cummings}
\newacronym{qLDPC}{qLDPC}{quantum low\hyp density parity\hyp check codes}
\newacronym{DFT}{DFT}{density\hyp functional theory}
\newacronym{DMFT}{DMFT}{dynamical mean\hyp field theory}
\newacronym{DMRG}{DMRG}{density\hyp matrix renormalization group}
\newacronym{1DMRG}{1DMRG}{single-site density\hyp matrix renormalization group}
\newacronym{2DMRG}{2DMRG}{two-site density\hyp matrix renormalization group}
\newacronym{DMRG3S}{DMRG3S}{strictly single-site density\hyp matrix renormalization group}
\newacronym{iDMRG}{iDMRG}{inifinite\hyp size density\hyp matrix renormalization group}
\newacronym{tDMRG}{tDMRG}{time\hyp dependent density\hyp matrix renormalization group}
\newacronym{QMC}{QMC}{quantum Monte Carlo}
\newacronym{AIM}{AIM}{Anderson impurity model}
\newacronym{SIAM}{SIAM}{single impurity Anderson model}
\newacronym{LDA}{LDA}{local\hyp density approximation}
\newacronym{LBNL}{LBNL}{Lawrence Berkeley National Laboratory}
\newacronym{VQE}{VQE}{variational\hyp quantum eigensolver}
\newacronym{ED}{ED}{exact diagonalization}
\newacronym{QPT}{QPT}{quantum phase transition}
\newacronym{QCP}{QCP}{quantum critical point}
\newacronym{ETH}{ETH}{eigenstate thermalization hypothesis}
\newacronym{EHM}{EHM}{extended Hubbard model}
\newacronym{BHW}{BHW}{Bose\hyp Hubbard wheel}
\newacronym{AKLT}{AKLT}{Affleck\hyp Lieb\hyp Kennedy\hyp Tasaki}
\newglossaryentry{QR}{name={QR},description={QR decomposition}}
\newacronym{TNS}{TNS}{tensor\hyp network state}
\newacronym{SM}{SM}{supplemental material}
\newacronym{NOO}{NOO}{natural orbital occupation}
\newacronym{NO}{NO}{natural orbital}
\newacronym{LRO}{LRO}{long\hyp range order}
\newacronym{qLRO}{qLRO}{quasi\hyp long\hyp range order}
\newacronym{SC}{SC}{superconducting}
\newacronym{SCQ}{SCQ}{superconducting qubit}
\newacronym{VBGS}{VBGS}{valence bond ground-state}
\newacronym{PEPS}{PEPS}{projected entangled pair\hyp states}
\newacronym{ALS}{ALS}{alternating least squares}
\newacronym{BdG}{BdG}{Bogoljubov de-Gennes}
\newacronym{TFIM}{TFI}{transverse field Ising model}
\newacronym{PP}{PP}{projected purification}
\newacronym{BEC}{BEC}{Bose\hyp Einstein condensate}
\newacronym{JWT}{JWT}{Jordan\hyp Wigner transformation}
\newacronym{NISQ}{NISQ}{noisy intermediate scale quantum}
\newacronym{NN}{NN}{nearest\hyp neighbor}
\newacronym{NNN}{NNN}{next\hyp nearest\hyp neighbor}
\newacronym{SPDM}{SPDM}{single\hyp particle density matrix} 
\newacronym{HCB}{HCB}{hardcore bosons}
\newacronym{SF}{SF}{spinless fermions}
\newacronym{cQED}{cQED}{cavity quantum electro\hyp dynamics}
\newacronym{SOP}{SOP}{symmetric operation point}%
\usepackage{tikz}
\usepackage{graphicx}
\begin{document}%
	\title{Logical qubits with geometrically enhanced energetic and parametric protection}%
	\author{Reja~H.~Wilke}%
	\email{reja.wilke@physik.uni-muenchen.de}
	\affiliation{Institute for Theoretical Physics, ETH Z\"urich, CH-8093 Zurich, Switzerland}
	\affiliation{Department of Physics, Arnold Sommerfeld Center for Theoretical Physics (ASC), Ludwig-Maximilians-Universit\"{a}t M\"{u}nchen, 80333 M\"{u}nchen, Germany}%
	\affiliation{Munich Center for Quantum Science and Technology (MCQST), Schellingstr. 4, D-80799 M\"{u}nchen, Germany}
	\author{Leonard~W.~Pingen}%
	\affiliation{Department of Physics, Arnold Sommerfeld Center for Theoretical Physics (ASC), Ludwig-Maximilians-Universit\"{a}t M\"{u}nchen, 80333 M\"{u}nchen, Germany}%
	\author{Thomas~K\"ohler}%
	\affiliation{Department of Physics and Astronomy, Uppsala University, Box 516, S-751 20 Uppsala, Sweden}%
	\affiliation{SUPA, Institute of Photonics and Quantum Sciences, Heriot-Watt University, Edinburgh EH14 4AS, United Kingdom}%
	\author{Sebastian~Paeckel}%
	\email{sebastian.paeckel@physik.uni-muenchen.de}
	\affiliation{Department of Physics, Arnold Sommerfeld Center for Theoretical Physics (ASC), Ludwig-Maximilians-Universit\"{a}t M\"{u}nchen, 80333 M\"{u}nchen, Germany}%
	\affiliation{Munich Center for Quantum Science and Technology (MCQST), Schellingstr. 4, D-80799 M\"{u}nchen, Germany}
	\date{\today}%
	\begin{abstract}%
	Breaking the error\hyp threshold marks a milestone in establishing quantum advantage for a wide range of relevant problems.
	One possible route is to encode information redundantly in a logical qubit by combining several noisy qubits, providing an increased robustness against external perturbations.
	We propose a setup for a logical qubit in a generalized Tavis-Cummings model consisting of physical qubits coupled to a logical all\hyp to\hyp one element.
	Our design is based on a recently discovered geometric stabilizing mechanism in the closely related~\gls{BHW}, which manifests as energetically well\hyp separated clusters of many\hyp body eigenstates.
	We show that even in the presence of typical fabrication uncertainties, their occurrence and separation is extremely robust.
	Introducing an additional physical qubit coupled to the central element yields replicas of these clusters, for which we introduce two prototypical coupling schemes.
	(i) A local coupling protocol exploits the spectral separation to encode information redundantly.
	(ii) A non\hyp local coupling protocol additionally generates a symmetry\hyp protected operating point at which virtual transitions produce no leading\hyp order differential energy shift within the logical manifold.
	Our results establish geometrically stabilized many\hyp body spectral clusters as tools to construct logical qubit manifolds with independently tunable energetic and parametric error protection.
	\end{abstract}%
	\glsresetall%
	\maketitle%
	%
	%
	\section{\label{sec:introduction}Introduction}%
	Quantum algorithms obtain polynomial and super\hyp polynomial speed\hyp ups compared to classical algorithms \cite{Grover1996,Shor1999} on a selected set of problems~\cite{Preskill2018}.
	During the past decades, this
	promise has given rise to the development of schemes that mitigate the effects of noise and errors, which typically set a time limit to store and process information in a physical system \cite{Unruh1995}.
	Importantly, it has been shown that qubit error rates below a certain threshold allow for arbitrarily accurate quantum computation \cite{Knill1998, Preskill1998, Aharonov2008}.
	Hence, lowering error rates below this threshold marks the central endeavor towards the practical application of quantum algorithms.
	To account for the ubiquitous presence of error sources, corrupting both the information represented by the qubit as well as its readout, error\hyp correction schemes, such as error mitigation~\cite{Li2017,Temme2017,QuantumErrorMitigation} or error\hyp correction codes~\cite{Calderbank1996,Steane1996,Dennis2002,Gottesman2010} have been introduced.
	These strategies share the underlying idea that information is represented redundantly and/or non\hyp locally, and can be recovered by repeated measurements or operations on an ensemble of qubits~\cite{Kitaev1997,Kitaev2003}.
	In particular, error\hyp correction codes distribute the quantum information non\hyp locally such that measurements allow for an active correction~\cite{Krinner2022,Takeda2022,Ryan-Anderson2022,QKAI2023}.

	The main obstacle is the introduced overhead, shifting the problem of implementing fault\hyp tolerant quantum computation to that of realizing quantum processors with a significantly larger number of physical qubits than operational, logical qubits.
	Here, remarkable progress has been achieved in addressing that problem very recently, for instance using reconfigurable neutral\hyp atom arrays, where high\hyp rate~\gls{qLDPC} have enabled up to $96$ logical qubits to be realized from only a few hundred physical atoms, reaching physical\hyp to\hyp logical ratios approaching $4$--$5$:$1$~\cite{Bluvstein2024,Zhou2025,Bluvstein2026}.
	Despite these rapid developments, the practically most relevant platforms remain based on \glspl{SCQ}, which continue to constitute the backbone of recent advances in fabricating large\hyp scale quantum processors~\cite{Vigneau2025,Acharya2025,Bravyi2024} and have brought forward new possibilities to compose logical qubits out of several~\gls{SCQ} elements, most notably through below\hyp threshold surface\hyp code memories and newly proposed \gls{qLDPC}\hyp based architectures that promise to cut the physical\hyp qubit overhead by up to an order of magnitude~\cite{DiVincenzo2009,Gambetta2017,Andersen2023,Bravyi2024}.
	Here, the main obstacles remain the comparatively high error rates combined with limited qubit connectivity, which still require on the order of $100$~\glspl{SCQ} per logical qubit in the best current experimental demonstrations, and continue to demand $100$--$1000$~\glspl{SCQ} in surface\hyp code\hyp based designs, even as \gls{qLDPC}\hyp based proposals target ratios closer to $10$--$50$:$1$~\cite{Acharya2025,Bravyi2024}.
	One route to reducing this overhead is to pursue hardware\hyp efficient logical qubits with intrinsic, passive protection, whether from the static Hamiltonian of the $0$\hyp$\pi$ qubit~\cite{Gyenis2021,Kolesnikow2025} or from dissipation\hyp engineered bosonic cat qubits~\cite{Putterman2025}.
	The former echoes the long\hyp standing principle of encoding information non\hyp locally into a symmetry\hyp protected, decoherence\hyp free subspace~\cite{ZanardiRasetti1997,DuanGuo1997,LidarChuangWhaley1998}.
	Collective enhancement effects, such as superradiance, offer another example of this strategy.
	Originally described by Dicke~\cite{Dicke1954}, superradiance arises when $L$ emitters couple to a shared field, resulting in a $\sqrt L$ enhancement of the collective interaction strength.
	This principle was first harnessed for quantum information applications in the \gls{DLCZ} protocol for quantum memory~\cite{Duan2001}. 
	Experimental realizations of such collective coupling, however, initially relied on atomic ensembles that behave, on average, as a single Tavis\hyp Cummings system, without resolving or addressing the individual atoms involved~\cite{Colombe2007,Brennecke2007}. %
	Site\hyp resolved control and readout, together with the higher connectivity available in modern quantum simulators, have sparked a revival of interest in superradiance. 
	Collective couplings have since been demonstrated on platforms including circuit QED~\cite{Du2026} and neutral\hyp atom arrays~\cite{Yan2023}.
	Here, we propose a logical qubit construction that exploits this collective enhancement by resonantly coupling $L$ stabilizer qubits to a controllable central element, see~\cref{fig:wheel_local} and~\cref{fig:wheel_nonlocal}.
	Our construction achieves high resilience against external perturbations using only a small number of qubits, $\mathcal O(10)$, coupled to a cavity mode as described by the \gls{TC} model~\cite{Tavis1968}, and is not tied to any specific experimental platform.
	The \gls{TC} model has been realized across multiple platforms, including~\gls{SCQ}~\cite{Fink2009} and neutral atoms~\cite{Liu2023}, which makes our proposal suitable for multiple hardware implementations.
	\gls{SCQ} processors with a star topology have already been demonstrated to function as quantum simulators~\cite{Renger2026,Leppakangas2025}.
	More specifically, our proposal is based on the clustering and separation of many\hyp body eigenstates of the~\gls{BHW}, illustrated in~\cref{fig:wheel_bosehubbard}, in the limit of infinitely strong repulsive interactions~\cite{Dongen1991,Vidal2011,Mate2021,Wilke23}.
	The separation of the energetically lowest cluster of many\hyp body eigenstates has been shown recently to be tunable either by the coupling $s$ between the wheel's ring sites and the center, or by increasing the wheel's coordination number $L$~\cite{Wilke23}.
	The resulting gap in the many\hyp body spectrum scales as $s\sqrt L$, yielding a geometric mechanism to separate out a cluster of many\hyp body eigenstates.
	These properties are inherited from a \textit{bright} single\hyp particle eigenstate with gap $\propto \sqrt L$.
	In order to create logical qubit states, we introduce an additional control qubit, for which we investigate two different coupling protocols.
	These generate distinctive sets of many\hyp body eigenstates in the low\hyp energy part of the spectrum which can be isolated via geometrically engineering the many\hyp body gap of the~\gls{BHW}.
	Because our subsequent analysis is based on these structured many\hyp body spectra, first, we rigorously show the robustness of the many-body spectrum against the impact of experimentally unavoidable imperfections in the coupling between the physical qubits and the central element.
	Based on this reslience, we demonstrate the emergence of a logical subspace that is energetically separated from the remaining eigenstates with a many\hyp body gap $\propto \Tilde{s} = s\sqrt{L}$.
	While these points in general demonstrate the energetic protection of the manifold of logical qubit\hyp states, the error\hyp resilience within this manifold depends on the specific coupling protocol.
	Here, we follow two different approaches by introducing (i) a \textit{local} coupling between the control qubit and the central element only, and (ii) a \textit{non\hyp local} coupling between the control qubit and the bright single\hyp particle state.
	The local protocol is cheap with respect to the number of gate operations, yet it requires an additional frequency detuning $\mu_c$ between the control and stabilizer qubit in order to generate two distinctive sets of many\hyp body eigenstates forming the logical $\ket{0}$ and $\ket{1}$ states.
	The non\hyp local protocol leverages the extensively scaling single\hyp particle gap of the bright mode in the vicinity of an avoided level\hyp crossing.
	This allows to construct logical $\ket{0}$ and $\ket{1}$ states that can be operated at~\gls{SOP}, for which differential energy shifts are suppressed to first order~\cite{Koch2007}.
	For both protocols, we study the realization of qubit operations.
	For the case of the \textit{locally} coupled logical qubit we investigate the single qubit $X$\hyp gate fidelity.
	We find that the readout error rate of the logical qubit can be decreased by more than an order of magnitude, compared to the single\hyp qubit case, when realizing the wheel\hyp geometry with $L = 20$ physical qubits, exploiting geometric stabilization.
	This yields a total single qubit\hyp gate fidelity of $F>0.999$.
	For the case of the \textit{non\hyp locally} coupled qubit, we show that coherent control of the qubit over the Bloch sphere can be achieved by tuning the external experimental parameters.
	The paper is organized as follows: 
	In \cref{sec:exp-proposal}, we introduce the~\gls{BHW} alongside it's properties and examine the effect of disorder on the ring\hyp to\hyp center hopping simulating experimental imperfections.
	In \cref{sec:logical-qubit}, we propose the experimental setup for a logical qubit.
	We introduce the coupling protocols and characterize the gate application and qubit readout.
	\section{\label{sec:exp-proposal}The Bose\hyp Hubbard wheel}%
	\begin{figure*}[t]
		\centering%
		\def\nmax{13}%
		\def\radius{1.1}%
		\def\qubitwidth{0.6em}
		\subfloat[\label{fig:wheel_bosehubbard}Bose\hyp Hubbard wheel]%
		{%
			\ifthenelse{\boolean{buildtikzpics}}%
			{%
				\input{figures/bose_hubbard_wheel}%
			}%
			{%
				\includegraphics{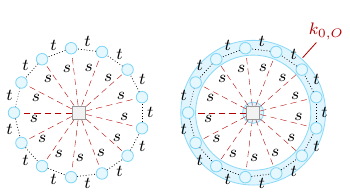}%
			}%
		}%
		\def\nnodes{7}%
		\def\qubitSeperation{0.4}%
		\def\manifoldLength{0.6}%
		\def\manifoldDistance{0.065}%
		\def\topManifold{0.7}%
		\def\middleManifold{-0.6}%
		\def\bottomManifold{-1.2}%
		\subfloat[\label{fig:wheel_local}Local coupling protocol]%
		{%
			\ifthenelse{\boolean{buildtikzpics}}%
			{%
				\input{figures/local_coupling_protocol}%
			}%
			{%
				\includegraphics{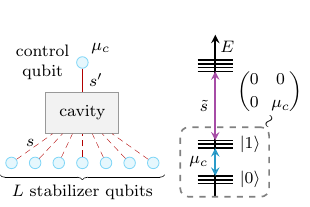}%
			}%
		}%
		\subfloat[\label{fig:wheel_nonlocal}Non\hyp local coupling protocol]%
		{%
			\ifthenelse{\boolean{buildtikzpics}}%
			{%
				\input{figures/non_local_coupling_protocol}%
			}%
			{%
				\includegraphics{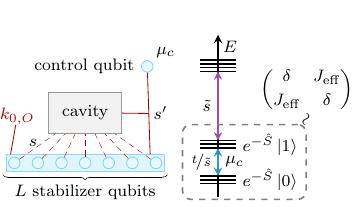}%
			}%
		}%
		\caption
		{
			\protect\subref{fig:wheel_bosehubbard} Lattice geometry of the Bose\hyp Hubbard wheel (left), see \cref{eq:hubb-wheel}, and visualization of the non\hyp trivial $k_0$\hyp mode (right).
			The $L$ ring sites, each of which is coupled to the center site with an amplitude $s$, exhibit nearest\hyp neighbor hopping with an amplitude $t$.
			The specific $k_0$\hyp mode is characterized by an equal superposition of single\hyp particle excitations on the ring, $k_{0,O}$, coupled to the center site.
			\protect\subref{fig:wheel_local} Illustration of the experimental setup based on the local coupling protocol discussed in~\cref{sec:logical-qubit:local}.
			In the proposed setup, the ring sites are realized as physical (stabilizer) qubits coupled to a controllable central element, here the cavity.
			In addition, we introduce a control qubit with on\hyp site potential $\mu_c$, exclusively coupled to the cavity with an amplitude $s^\prime$.
			The control qubit allows for read out and manipulation of logical information.
			The logical qubit gap is controlled by the on\hyp site potential $\mu_c$, the energetic seperation from the higher\hyp energy states scales with $\tilde s$, see \protect\subref{fig:wheel_local} (right).
			\protect\subref{fig:wheel_nonlocal} Illustration of the experimental setup based on the non\hyp local coupling protocol discussed in~\cref{sec:logical-qubit:non-local}.
			The logical qubit Hamiltonian is controlled by the ratio $t/\tilde s$ and the on\hyp site potential $\mu_c$, the energetic seperation from the higher\hyp energy states scales with $\tilde s$, see \protect\subref{fig:wheel_nonlocal} (right).
			For the local protocol, coupling between the stabilizer qubits is optional, while the non\hyp local protocol exploits $t$ to implement coherent logical\hyp qubit control~\cref{eq:wheel+probe:non-local:logical-qubit}.
		}
	\end{figure*}

	We consider a Bose\hyp Hubbard model on a wheel geometry in the limit of large interactions $U \rightarrow\infty$, where the bosons become hard\hyp core \cite{Wilke23,Vidal2011,Mate2021}, which will be referred to as~\gls{BHW} in the following.
	The Hamiltonian of the system is given by (see~\cref{fig:wheel_bosehubbard})
	\begin{align}%
		\hat H_\mathrm{wheel}%
		\equiv%
		-t \sum_{j=0}^{L-1} \left( \hat h^\dagger_j \hat h^\nodagger_{j+1} + \mathrm{h.c.} \right) -%
		\sum_{j=0}^{L-1} \left( s^\nodagger_j\hat h^\dagger_j \hat h^\nodagger_\odot + \mathrm{h.c.} \right) \;. %
		\label{eq:hubb-wheel}%
	\end{align}%
	Here, $t$ denotes the hopping on the outer ring of the wheel, consisting of $L$ sites, and $s_j = s \mathrm e^{\mathrm i k_0 j}$ describes a $k_0$\hyp modulated ring\hyp to\hyp center hopping.
	 $\hat h^{(\dagger)}_j$ corresponds to the \gls{HCB} annihilation (creation) operator on the $j$\hyp th site of the ring.
	The index $\odot$ denotes the respective operators on the center site or all\hyp to \hyp one element, which we will refer to as the cavity in the following.
	We consider periodic boundary conditions $\hat h^{(\dagger)}_L \equiv \hat h^{(\dagger)}_0$.
	In the limit $\frac{s}{t}\rightarrow 0$ (ring geometry), the system exhibits a quasi\hyp\gls{BEC} with ground\hyp state occupation $\propto \sqrt{N}$, where $N$ denotes the number of \gls{HCB} in the system \cite{Rigol2004, Rigol2005, Lieb1963a, Lieb1963}.
	In the opposite limit $\frac{s}{t}\rightarrow \infty$ (star geometry)\cite{Dongen1991}, where the system coincides with the~\gls{TC} model, the system exhibits a true \gls{BEC} with ground\hyp state occupation $\propto N$ \cite{Tennie2017}.
	In our recent work \cite{Wilke23}, we solved the full many\hyp body problem of the \gls{BHW} by mapping the system to a periodic ladder of spinless fermions using a Jordan\hyp Wigner transformation~\cite{Jordan1928}.
	\begin{table}%
		\centering%
		\caption%
		{%
			\label{tab:occupations}%
			Possible occupations $n_{k_0}$ of the $k_0$\hyp mode and the properties of the corresponding states.
			The table provides an overview of the parities $\pi(n_{k_0})$, presence and scaling of the gap as well as the presence of a \gls{BEC}.
			The properties of the $k_0$\hyp subspace are passed on to the many\hyp body\hyp spectrum.
		}%
		\begin{tabular}{ccccc}%
			\toprule%
			$n_{k_0}$ & $0$ & $1_+$ & $1_-$ & $2$ \\%
			\midrule%
			$\pi(n_{k_0})$ & even& odd & odd & even \\%
			\midrule%
			gap & $\times$& $\propto\Tilde{s}$ & $\propto-\Tilde{s}$ & $\times$ \\%
			\midrule%
			BEC & $\times$ & $\times$ & $\checkmark$ & $\times$ \\%
			\midrule%
			type & dark & bright & bright & dark\\%
			\bottomrule%
		\end{tabular}%
	\end{table}%
	We showed that the many\hyp body spectrum is characterized by the formation of branches containing clusters of eigenstates.
	These branches exhibit a distinct scaling behavior, a feature that is inherited from the single\hyp particle spectrum.
	In particular, it contains a distinct single\hyp particle eigenstate, the $k_0$\hyp mode, which, if singly occupied, gives rise to the existence of two marginal single\hyp particle eigenstates with energies $E_\pm$.
	Note that this distinct single\hyp particle eigenstate corresponds to a symmetric superposition of the ring sites coupled to the center site, see right panel of~\cref{fig:wheel_bosehubbard}, and is thus the bright, collectively-coupled state familiar from Dicke's treatment of symmetric superpositions~\cite{Dicke1954}, which would exhibit superradiant (collectively enhanced) decay if coupled to a dissipative channel.

	Due to it's distinct properties, the occupation $n_{k_0}$ of the $k_0$\hyp mode controls the scaling of the many\hyp body branches:
	There are two trivial states with $n_{k_0}=0,2$ (dark states), which give rise to many\hyp body eigenstates with energies of the order of the band width $t$.
	Furthermore, there are the two marginal single\hyp particle states, which we label by $n_{k_0}=1_\pm$ (bright states).
	These generate a cluster of non\hyp trivial (i.e., featuring a~\gls{BEC}) many\hyp body eigenstates, which energetically separate from the trivial states $\propto \pm s \sqrt L \equiv \pm \Tilde s$.
	Due to the scaling $\propto \sqrt L$, this separation of many\hyp body eigenstates can be understood as a geometric stabilizing mechanism.
	From the remaining single\hyp particle eigenstates with $k\neq k_0$, we construct a basis of the many\hyp body Hilbert space in terms of Slater determinants denoted by $\ket{\mathrm{FS}_N} = \ket{\left\{n_k \right\}_{k\neq k_0}}$, with $n_k \in \left\{0,1 \right\}$ and $N$ particles.
	The full many-body eigenstates of the~\gls{BHW} are then given by
	\begin{align}
		&\ket{E_{n_{k_0} = 0}(\mathrm{FS}_{N})} = \ket{0}\ket{FS_{N}}\;, \label{eq:hubb-wheel:estate:0}\\
		&\ket{E_{n_{k_0} = \pm}(\mathrm{FS}_{N-1})} = \ket{\pm}\ket{FS_{N-1}}\;, \label{eq:hubb-wheel:estate:pm}\\
		&\ket{E_{n_{k_0} = 2}(\mathrm{FS}_{N-2})} = \ket{2}\ket{FS_{N-2}}\;, \label{eq:hubb-wheel:estate:2}
	\end{align}
	where the single\hyp particle states $\ket{\pm}$ diagonalize the $n_{k_0} = 1_{\pm}$ subspace.
	For each $L,N$, $\mathcal N_{n_{k_0}}(L,N) = \binom{L}{N-n_{k_0}}$ Slater determinants can be constructed, and their number determines the amount of many\hyp body states in each branch.
	We note that the many-body eigenstates obtained from the solution strategy of the~\gls{BHW} agree with those of the single\hyp photon~\gls{TC} model~\cite{Tavis1968} in the limit $t\rightarrow 0$ and thus provide an extension of the many-body solution of the~\gls{TC} model to the case of qubit-qubit interactions.\\
	\textit{Presence of noise.--}
	Errors in the realization of the \gls{BHW} can be classified into unitary and non\hyp unitary ones:
	To the latter belong for instance the decay of physical qubit states, dephasing or bitflip errors, which we will address later.
	Unitary errors are generated by an imperfect realization of the \gls{BHW} Hamiltonian.
	For the case of interactions on the outer ring, the robustness has been demonstrated previously using~\gls{DMRG}:
	The \gls{BEC} exists even in the presence of interactions on the outer ring as a direct consequence of the non\hyp local coupling between the $k_0$\hyp mode and the center site~\cite{Wilke23}.
	Here, we show that the stabilizing mechanism of the \gls{BHW} -- which the stability of the proposed logical qubit relies on -- remains intact despite experimental imperfections, which translate into perturbations of the wheel\hyp to\hyp center couplings $s_j$.
	To model these imperfections, we consider the effect of perturbations $\delta s_j$ to the ring\hyp to\hyp center hopping amplitudes $s_j = s \mathrm e^{\mathrm i k_0 j}
		\rightarrow (s + \delta s_j) \mathrm e^{\mathrm i k_0 j}$.
	The perturbations $\delta s_j$ are considered to be normal distributed, independent, random variables, with standard deviation $\sigma$.
	Bounds on the induced shifts of the single\hyp particle spectrum can be derived and, therefrom, the robustness of the many\hyp body gap, see~\cite{suppMat}.
	In particular, for the marginal single\hyp particle eigenvalues $E_\pm$, the eigenvalue equation can be reformulated as a self\hyp consistent problem in terms of the perturbations $\delta s_j$.
	Expanding the self\hyp consistency equation to first order in the perturbations $\delta s_j$, the extremal solutions to this eigenvalue problem, $\tilde{E}_\pm$, can be bounded.
	An average over the normal distributed perturbations reproduces the unperturbed marginal energies $E_\pm$~\cite{Wilke23} up to second order corrections in $\sigma$,
	\begin{figure}[t]%
		\centering%
		\ifthenelse{\boolean{buildtikzpics}}%
		{%
			\input{figures/noisy_single_particle_spectrum_estimation}
	  	}
  		{
  			\includegraphics{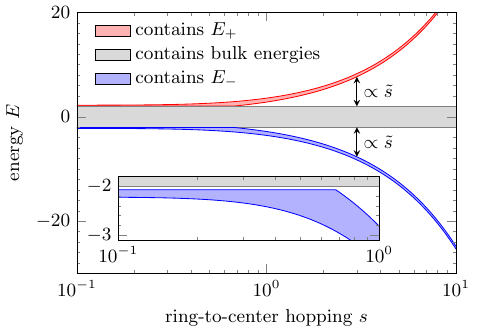}%
  		}
		\caption{%
			\label{fig:wheel:disorder:sp:eigenvalues}
			Demonstration of the established bounds for the single-particle spectrum of the perturbed~\gls{BHW}.
			The data is generated for $N = 1$ particle in a system with $L=6$ ring sites, $k_0 = \pi / 3$ and normal distributed perturbation $\delta s_j$ with standard deviation $\sigma/t = 0.4$.
			The marginal single\hyp particle energy $E_+$ ($E_-$) is contained in the red (blue) shaded area, which separates $\propto \Tilde{s}$, where $\Tilde{s}$ denotes the re\hyp scaled hopping amplitude.
			All remaining energies are confined to the grey bulk, i.e., to $]-2t, 2t[$, specified by the hopping on the outer ring.
		}%
	\end{figure}%
	\begin{equation}
		\begin{aligned}
		\mathrm{E} \left[ \Tilde{E}_\pm \right] &= -t \cos{k_0} \pm \sqrt{(t \cos{k_0})^2 + \Tilde{s}^2} + \mathcal{O}(\sigma^2/s)\\ &\equiv E_\pm + \mathcal{O}(\sigma^2/s)\; .
		\label{eq:wheel:disorder:Etilde}
		\end{aligned}
	\end{equation}
	Therefore, the crucial property of the~\gls{BHW}, i.e., the separation of two single\hyp particle eigenstates $\propto \pm \Tilde{s}$ is also robust against small, random perturbations of the ring\hyp to\hyp center hopping $s$.
	For an example realization of the noisy~\gls{BHW}, the single\hyp particle spectrum is shown in~\cref{fig:wheel:disorder:sp:eigenvalues}.
	The derived bounds on the marginal energies $E_\pm$, see~\cite{suppMat}, span the blue and red shaded regions separating from the bulk spectrum $\propto \tilde s$ and demonstrate the stability of the single\hyp particle gap against perturbations, here at the example of a disorder realization with standard deviation $\sigma/t=0.4$.
	By treating $\delta s_j$ perturbatively, we show that the extensively scaling gap in the many\hyp body spectrum between trivial and~\gls{BEC} states is robust also in the presence of random perturbations $\delta s_j$ of the ring\hyp to\hyp center hopping, as long as $\delta s_j \ll \Tilde s$, see~\cite{suppMat}.
	We want to stress that this result is far from being trivial because the number of perturbations $\delta s_j$ of the system's Hamiltonian scales with the number of lattice sites $L$ on the outer ring.
	However, the robustness can be understood by noting that under the Jordan\hyp Wigner transformation, many\hyp particle eigenstates of the~\gls{BHW} \cref{eq:hubb-wheel} are described by Slater determinants $\ket{FS_\mathrm{N}} = \ket{\left\{n_k \right\}_{k\neq k_0}}$ of single\hyp particle modes ~\cite{Wilke23}.
	In the presence of disorder, these Slater determinants are constructed from single\hyp particle eigenstates whose energies are shifted by random perturbations $E \rightarrow E + \delta E$.
	In leading order, the distribution of the $\delta E$ is dominated by the normal distributed perturbations $\delta s_j$.
	The total contribution of these perturbed Slater determinants to the many\hyp body energies is obtained by summing over all occupied single\hyp particle states, which effectively constitutes an average over the random perturbations $\delta E$ and, thus, the perturbations average out with standard deviation $\sim \sigma$.
	Importantly, this conclusion also applies to random variations of the stabilizer qubit frequencies, and our results can therefore be used to estimate the acceptable imperfections in these frequencies such that, in an actual practical realization, the condition $\delta s_j \ll \Tilde s$ can be satisfied.
	The scaling and the resulting stability of the many\hyp body gap against interactions and disorder render the system a promising candidate for reliably storing (logical) information in the energetically separated subspace.
	As a brief side remark, it should be noted that this observed robustness suggests a possible route to realizing a~\gls{BEC} in quantum simulators consisting of qubits coupled to a high\hyp connectivity central element.	%
	Previous works have shown the presence of a~\gls{BEC} in an approximate \gls{TC} models realized via thermal atom ensembles~\cite{Colombe2007,Brennecke2007}.
	\section{\label{sec:logical-qubit} The Bose\hyp Hubbard wheel with a control qubit}
	\begin{figure*}[t]
		\centering
		\subfloat[\label{fig:many_body_spectrum}]%
		{%
			 \ifthenelse{\boolean{buildtikzpics}}%
			{%
				\input{figures/entire_spectrum}%
			}%
			{%
				 \includegraphics{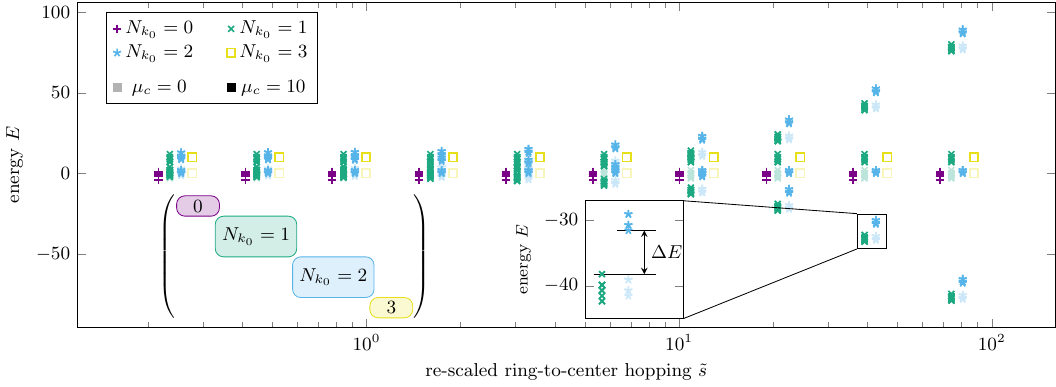}%
			}%
		}
		\subfloat[\label{fig:many_body_spectrum_occupancies}]%
		{%
			\ifthenelse{\boolean{buildtikzpics}}%
			{
				\input{figures/activate_probe_site}
			}%
			{%
				\includegraphics{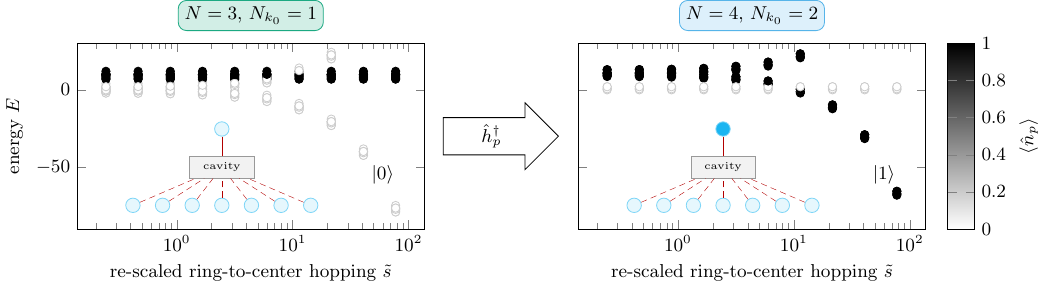}%
			}%
		}
		\caption%
		{Many\hyp body spectrum of $\hat H_{\mathrm{qubit}}$,~\cref{eq:wheel+probe:hamiltonian:local}, as a function of the re\hyp scaled hopping amplitude $\Tilde{s}$.
			\protect\subref{fig:many_body_spectrum} Many\hyp body spectrum highlighting the $N_{k_0}$\hyp sectors involved in the construction of the logical qubit.
			The color and shape of the data points indicate the $N_{k_0}$\hyp sector of the respective states, see left inset.
			The data points of different $N_{k_0}$ are plotted next to each other and using distinct shapes, resulting in batches of four corresponding to a single value of $\Tilde{s}$.
			Different data points for a given $N_{k_0}$ within such a batch correspond to different Slater determinants $\ket{\mathrm{FS}_{N-N_{k_0}}}$.
			We show the spectra for two parameter configurations, differing only in the choice of $\mu_c$, on top of each other: 
			High (low) opacity data points correspond to $\mu_c = 10$ ($\mu_c = 0$), visualizing the emerging two\hyp level system at $\mu_c > 0$, where the $N_{k_0} = 2$ sector is shifted w.r.t the $N_{k_0} = 1$ sector.
			The right inset indicates the ground $(\ket{0})$ and excited $(\ket{1})$ states of the logical qubit where $\Delta E$ suffices as a lower bound for the gap, see~\cref{eq:gap}.
			\protect\subref{fig:many_body_spectrum_occupancies} shows the process of exciting the logical qubit $\ket{0} \rightarrow \ket{1}$.
			The color coding denotes the expected control\hyp qubit occupation $\braket{\hat n_c}$.
			The left plot shows the $N_{k_0} = 1$ sector of \protect\subref{fig:many_body_spectrum}, characterized by $\braket{\hat n_c} = 0$ and indicated by triangular shapes, which constitutes the ground state of the logical qubit.
			Acting with $\hat h^\dagger_c$ performs an excitation of the logical qubit.
			The excitation leads to the $N_{k_0} = 2$ sector, characterized by $\braket{\hat n_c} = 1$ and indicated by circles, of a system with $N+1$ particles, see right plot, constituting the excited state of the logical qubit.
			The readout accuracy can be controlled by tuning $\nicefrac{s^\prime}{\Tilde s}$.
			While this behavior can be observed for any $L,N$, in this particular instance, we consider $L=6$, $N=3$ as well as control\hyp qubit\hyp to\hyp center hopping $s^\prime = 0.01$.
			}	
		\label{fig:wheel:probe:spec}
	\end{figure*}
	The robustness of geometric stabilization against perturbations provides a novel perspective on error\hyp resilient logical qubits.
	To this end, an additional control qubit is introduced doubling the number of many\hyp body eigenstates.
	The central idea is now to leverage the energetic separation of the clusters of many\hyp body eigenstates of the~\gls{BHW} and form logical subspaces via additional couplings between the control qubit and the cental element, see~\cref{fig:wheel_local}.
	In general, there are two conceptually different approaches to achieve this goal: The introduction of an external field (see~\cref{sec:logical-qubit:local}) or via the hybridization of the energetically separated clusters via non\hyp local interactions (see~\cref{sec:logical-qubit:non-local}).
	Note that both protocols exhibit a similar structure of the many\hyp body spectrum: A low\hyp energy subspace is separated by a many\hyp body gap, which scales $\propto \Tilde{s}$ introducing an energetic protection of states within this subspace against thermal fluctuation.
	We will introduce and discuss realizations of both approaches in the following and demonstrate the improvements gained by the energetic protection at the example the local coupling protocol.
	\subsection{\label{sec:logical-qubit:local} Local coupling protocol}
	A straight forward procedure to lift the degeneracy between the many\hyp body eigenstates of the~\gls{BHW} in the energetically low\hyp lying cluster is to introduce an effective onsite potential $\mu_c$ that acts on the control qubit only, which can be realized by a frequency detuning between the stabilizer qubits and the control qubit.
	We then allow for a weak hybridization $s^\prime \ll \tilde s$ between the control qubit and the cavity yielding the Hamiltonian
	\begin{equation}
		\hat H_\mathrm{qubit}
		=
		\hat H_\mathrm{wheel}
		+
		s^\prime \left(\hat h^\dagger_\odot \hat h^\nodagger_c + \mathrm{h.c.} \right)
		+
		\mu^\nodagger_c \hat n^\nodagger_c \;, \label{eq:wheel+probe:hamiltonian:local}
	\end{equation}
	where $\hat h^{(\dagger)}_c$ denotes annihilation (creation) operator for the additional control qubit, $\hat n^\nodagger_c = \hat h^\dagger_c \hat h^\nodagger_c$ the occupation of the control qubit, and $\mu^\nodagger_c$ a chemical potential.
	In the following we consider the general case of a finite hopping amplitude $t=1$, unless stated differently.
	The width of the clusters in the many\hyp body spectrum is given by $2t$~\cite{Wilke23}.
	Upon introducing the control qubit, the many\hyp body branches of the \gls{BHW} are being duplicated, owing to the two\hyp dimensional local Hilbert space of the control site.
	The energetically low\hyp lying many\hyp body eigenstates can still be separated from the remaining spectrum by geometric stabilization, i.e. increasing $\tilde s$.
	Notably, these energy clusters realize~\glspl{BEC} that can be distinguished by their constituting Slater determinants~\cite{Wilke23}.
	Thus, local and non\hyp hermitian perturbations, acting on the $L$ stabilizer qubits (the wheel's outer ring) couple many\hyp body eigenstates within the same cluster.
	This gives rise to a redundancy of the represented logical qubit state, which scales by the number of Slater determinants $\mathcal N_{n_{k_0}}(L,N)$ of the~\gls{BHW}.
	We make these statements more concrete in the following and show how they give rise to the desired properties of a logical qubit.
	For the solution of the many\hyp body problem of~\cref{eq:wheel+probe:hamiltonian:local} we introduce $\hat N_{k_0} = \hat n_c + \hat n_{k_0}$, the sum of the occupation of the control qubit and the $k_0$\hyp mode, which is a conserved quantity.
	Recall that any given state with $\hat n_{k_0} = 1_{-}$ can be identified with the non\hyp trivial, i.e. BEC, branch in the many-body spectrum of the \gls{BHW} (see~\cref{tab:occupations}).
	From the conservation of $\hat N_{k_0}$ it follows that for a given Slater determinant $\ket{\mathrm{FS}_N}$, $\hat H_{\mathrm{qubit}}$, decomposes into a block\hyp diagonal representation $\braket{\mathrm{FS}_N|\hat H_\mathrm{qubit} | \mathrm{FS}_N} = \bigoplus_{N_{k_0}} \hat h(\mathrm{FS}_N, N_{k_0})$.
	The individual block dimensions are given by
	\begin{equation}
		d_{N_{k_0}} = \operatorname{dim}[\hat h(\mathrm{FS}_N, N_{k_0})]
		=
		\begin{cases}
			1, &\text{if $N_{k_0} = 0,3$,} \\
			3, &\text{if $N_{k_0} = 1,2$}.
		\end{cases}
	\end{equation}
	The remaining eigenvalue problem of~\cref{eq:wheel+probe:hamiltonian:local} can be solved in each block, individually, by making use of the solution strategy for the \gls{BHW} reviewed in~\cref{sec:exp-proposal}.
	To explicitly account for the conservation of the overall particle number $N$, we re-sort the blocks, i.e., we fix  $N$ and obtain a block for each $N_{k_0}$ with $N-N_{k_0}$ occupied modes in the Slater determinants, see inset of \cref{fig:many_body_spectrum}.
	In this basis, eigenstates can be represented by
		\begin{equation}
		\ket{E_\nu(N, N_{k_0})}
		=
		\ket{\nu_{N_{k_0}}} \otimes \ket{\mathrm{FS}_{N-N_{k_0}}} \; ,
		\label{eq:wheel:probe:mbstates}
	\end{equation}
	where $\nu_{N_{k_0}} = 0,\dots, d_{N_{k_0}}-1$ labels the eigenstates in the corresponding $N_{k_0}$\hyp sectors sorted by their energies.
	While the full solution strategy for the eigenvalue problem of~\cref{eq:wheel+probe:hamiltonian:local} is given in~\cite{suppMat}, here, we will focus on the part of the many\hyp body spectrum relevant for the logical qubit.
\subsubsection{Logical qubit and $X$\hyp gate implementation}
	\begin{figure*}%
		\centering%
		\subfloat[\label{fig:error_rate:mu12}]%
		{%
			\def\tvalue{1}%
			\def\sprime{0.01}%
			\def\muValue{12.0}%
			\ifthenelse{\boolean{buildtikzpics}}%
			{%
				\def\xmin{0.9}%
				\def\xmax{10.1}%
				\def\ymin{1e-6}%
				\def\ymax{1.1}%
				\def\Lmin{6}%
				\def\Lmax{28}%
				\input{figures/error_rate}%
			}%
			{%
				\includegraphics{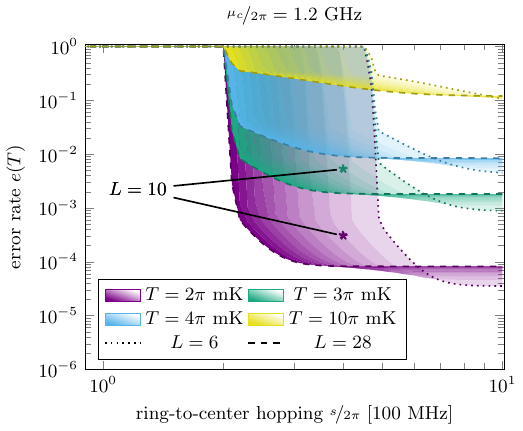}%
			}%
		}%
		\hfill
		\subfloat[\label{fig:error_rate:mu17}]%
		{%
			\def\tvalue{1}%
			\def\sprime{0.01}%
			\def\muValue{17.0}%
			\ifthenelse{\boolean{buildtikzpics}}%
			{%
				\def\xmin{0.9}%
				\def\xmax{10.1}%
				\def\ymin{1e-6}%
				\def\ymax{1.1}%
				\def\Lmin{6}%
				\def\Lmax{28}%
				\input{figures/error_rate}%
			}%
			{%
				\includegraphics{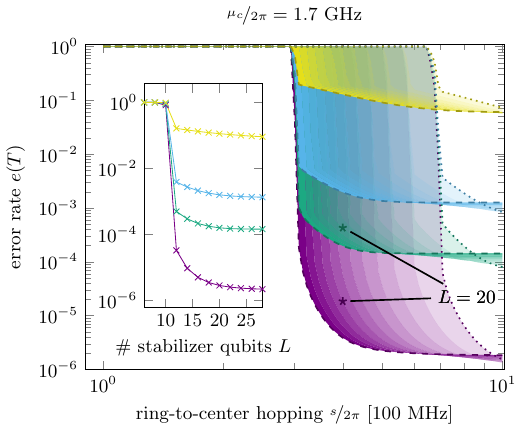}%
			}%
		}%
		\caption%
		{%
			\label{fig:wheel:probe:hot_wheels}
			Error rate $e(T)$ as a function of the ring\hyp to\hyp center hopping amplitude $s$ for different temperatures $T = 2\pi, 3\pi, 4\pi, 10\pi\,\mathrm{mK}$ and $\mu_c/2\pi = 1.2,1.7\, \mathrm{GHz}$ at fixed control site\hyp to\hyp center hopping $s^\prime/2\pi = 1\, \mathrm{MHz}$.
			The shaded areas are spanned by the values of $e(T)$ at fixed $T$ for different numbers of stabilizer qubits $L = 6,\ldots, 28$.
			The smallest (largest) system sizes $L = 6 (28)$ are denoted by the dashed (dotted) lines and the fill color saturation indicates the intermediate sizes.
			For $s<s_c (T,L)$, where $s_c (T,L)$ denotes a temperature\hyp{} and system\hyp size dependent critical value of $s$, increasing the amount of stabilizer qubits $L$ results in a lower error rate where the improvement becomes significant in the low\hyp temperature regime $T<10\pi\,\mathrm{mK}$.
			For $s>s_c (T)$, this behavior is reversed and adding more stabilizer qubits does not improve the error rate.
			The effect of geometric stabilization is to reduce the coupling strength $s$ between stabilizer qubits and central mode, in order to achieve a certain error rate.
			This is illustrated in the inset, where we show the error rate as a function of $L$ for a fixed value of $s/2\pi=500\, \mathrm{MHz}$, which is at the upper limit of practically doable couplings~\cite{Yoshihara2017}.
 		}%
	\end{figure*}
	To realize an effective two\hyp level system, there are two symmetry sectors of particular interest, namely $\hat N_{k_0} = 1$ and $\hat N_{k_0} = 2$ with block dimension $d_{N_{k_0}} = 3$, each.
	To understand the effect of the control qubit on the unperturbed wheel's eigenstates we can treat $s^\prime / \Tilde s$ as a small perturbation.
	This is motivated by the fact that in practical realizations $\Tilde s$ should be as large as possible to stabilize the non\hyp trivial~\gls{BEC} phase, where the logical qubit lives.
	For $\hat H_{\mathrm{qubit}}$, the structure of the many\hyp body spectrum of the unperturbed wheel,$\hat H_{\mathrm{wheel}}$, is reproduced in the two relevant $N_{k_0}$\hyp sectors with corrections $\sim \mathcal O((\nicefrac{s^\prime}{\Tilde s})^2)$.
	An example for the many\hyp body spectrum as a function of $\Tilde s$ is shown in~\cref{fig:many_body_spectrum} for $s^\prime = 1$.
	\cref{fig:many_body_spectrum_occupancies} shows the control qubit occupation indicated by the fill color in each many\hyp body eigenstate of the $N_{k_0}=1$ (left) and $N_{k_0}=2$ (right) sector.
	Evaluating the occupation of the control qubit for the $N_{k_0} = 0,1$ eigenstates in the lower branch of the spectrum using perturbation theory, yields
	\begin{equation}
		\braket{\hat n_c}_{N_{k_0}=1} = \mathcal O \left(\left(\frac{s^\prime}{\Tilde s}\right)^2 \right)\;, \\
		\braket{\hat n_c}_{N_{k_0}=2} = 1 - \mathcal O \left(\left(\frac{s^\prime}{\Tilde s}\right)^2 \right) \; .
	\end{equation}
	Note that increasing $\Tilde s$, the occupations quickly saturate to either $\braket{\hat n_c} = 0$ or $\braket{\hat n_c} = 1$, depending on the respective symmetry sector.
	This defines an effective two\hyp level system between states in the two $N_{k_0}$ clusters
	\begin{equation}
		\ket{0} \stackrel{\sim}{=} \ket{\nu_{N_{k_0}=1}} \;, \quad
		\ket{1} \stackrel{\sim}{=}\ket{\nu_{N_{k_0}=2}}\;,
	\end{equation}
	for which the readout accuracy can be controlled by tuning $s^\prime/\Tilde s$.
	 In the limit $t = 0$, i.e., the star geometry, the many\hyp body energies corresponding to different configurations within a given $N_{k_0}$\hyp sector reduce to a single value, which allows us to define the energy gap of the logical qubit as
	 \begin{equation}
		\label{eq:gap}
	 	\Delta E = E_2 - E_1 \propto \mu_c \; ,
	 \end{equation}
	with
	\begin{align}
		E_1 =  E_{0}(N,1) \quad  \text{and} \quad
		E_2 =  E_{0}(N+1,2) \; .
	\end{align}
	This gap can be tuned via the chemical potential $\mu_c$ on the control qubit.
	Importantly, $\mu_c$ has no effect on the corrections of the control\hyp site occupations such that it can be treated as a free parameter that can be chosen as large as experimentally possible.
	For a fixed coupling $s^\prime/t$, we indeed find a linear relation between $\Delta E$ and the chemical potential $\mu_c$, see \cite{suppMat}.
	In our specific setup, $\mu_c$ can be controlled by the frequency detuning of the control qubit compared to the stabilizer qubits.
	Alongside $\tilde{s}$, $\mu_c$ is the relevant experimental parameter, both of which provide tuning knobs for the quality of the logical qubit.
	An important consequence of the proposed logical qubit is that logical gate operations only need to act on the control qubit and not on the stabilizer qubits.
	Thus, the frequencies of the stabilizer qubits may be increased beyond the logically addressable regime, allowing even larger detunings between the stabilizer qubits and the control qubit.
	For instance, typical frequencies for addressable transmon qubits range between $1\,\mathrm{GHz} - 10\,\mathrm{GHz}$, but this range can be increased to $100\,\mathrm{GHz}$ if no gate operations are required.
	This should be compared to anharmonicities in current transmons, which are of the order of $100\,\mathrm{MHz}$~\cite{WalraffcQED,Roth2023}.
	We now study the effect of finite\hyp temperatures on the measurement fidelity to distinguish the two logical states and the dependency on the different parameters $t, s, \mu_s$ and $s^\prime$.
	The thermal density operator of the ground state $\ket{0}$ is given by
	\begin{align}
		\hat \rho_0 (T) = \frac{1}{Z_\beta} e^{-\beta \hat H_\mathrm{qubit}}\; ,
		\label{eq:thermal_density}
	\end{align}
	where $\beta = 1/k_BT$, $k_B$ denotes the Boltzmann constant, and $Z_\beta = \operatorname{Tr} e^{-\beta \hat H_\mathrm{qubit}}$ the partition function.
	We define the fidelity of the $X$\hyp gate as the probability of finding the control qubit in a $\ket{1}$ state after performing the excitation
	\begin{equation}
		\hat \rho_0 (T) \mapsto  \hat h^\dagger_c \hat \rho_0 (T) \hat h^\nodagger_c = \hat \rho_1 (T) \; ,
	\end{equation}
	as function of T.
	This can be evaluated via the partial trace over the excited state projected into the lower branch of the $N_{k_0} = 2$ sector
	\begin{align}
		F(T,N) \equiv \sum_{\left\lbrace \mathrm{FS}_{N-2}\right\rbrace } \bra{E_0(N+1, 2)} \hat \rho_1 (T) \ket{E_0(N+1, 2)}\;,
		\label{eq:wheel+probe:fideliT}
	\end{align}
	where the sum is over all Slater determinants with $N-2$ modes occupied.
	In the following, we choose a practically relevant energy scale, i.e., the upper edge of the strong\hyp coupling regime for couple transmon qubits $t/2\pi=100\,\mathrm{MHz}$ as unit of energy.
	From the fidelity, we immediately get the error rate for switching the state of a single qubit, which is of fundamental importance.
	It constitutes a lower bound for the error rate of two\hyp site gate operations and controls the approximation quality of quantum algorithms.
	We define $e(T)$ as the probability of finding our qubit in an eigenstate, which is not in the targeted lower $N_{k_0}=2$ branch: $e(T) = 1-F(T)$.
	In \cref{fig:wheel:probe:hot_wheels} we show $e(T)$ for different temperatures at half filling $N = L/2$ as a function of $s$ and detunings $\mu_c/2\pi = 1.2, 1.7\, \mathrm{GHz}$.
	The shaded areas are spanned by the values of $e(T)$ for different numbers of stabilizer qubits $L = 6,\dots, 28$, where the dashed (dotted) boundaries correspond to $L=6(28)$ and the fill color saturates with increasing $L$.
	For a fixed temperature and detuning, the dominating control parameter of the error rate is $s$ and we observe a steep decrease towards a lower bound set by the ratio $\mu_c/T$.
	The origin of this sharp drop is the separation of the lower $N_{k_0}=1,2$ clusters from the central clusters in the many\hyp body spectrum and signals the separation of the two logical qubit states.
	The width of the shaded areas is controlled by the relation between that separation and the rescaled coupling $\Tilde s = s \sqrt L$.
	For sufficiently large values of $s$, details of the many\hyp body spectrum such as the number of eigenstates per $N_{k_0}$\hyp sector become relevant.
	The onset of this regime of the ring\hyp to\hyp center hopping is indicated by the crossing points in the shaded areas, $s = s_c(T)$.
	As long as $s<s_c(T)$, the error rate can be reduced significantly by increasing the wheel's coordination number $L$, in particular at low temperatures $T < 10\pi\,\mathrm{mK}$.
	In this case, the crossing point $s_c(T)/2\pi \sim 1\,\mathrm{GHz}$ would be hard, if not impossible, to reach in experiment.
	Here, the importance of the wheel geometry becomes apparent: In order to achieve large fidelities (small error rates), the geometric stabilizing mechanism of the~\gls{BHW} with an extensively scaling many\hyp body gap is the key feature.
	Within the presented calculations single qubit\hyp gate fidelities of $F>0.999$ can be reached by coupling $L=20$ qubits to a cavity at a temperature of $3\pi \approx 10\,\mathrm{mK}$.
\subsubsection{Measuring the qubit state}
	\begin{figure}
		\centering%
		\subfloat[\label{fig:fidelityL10}]%
		{%
			\ifthenelse{\boolean{buildtikzpics}}%
			{%
				\def\Lvalue{10}%
				\def\muValue{12}
				\input{figures/fidelity}%
			}%
			{%
				\includegraphics{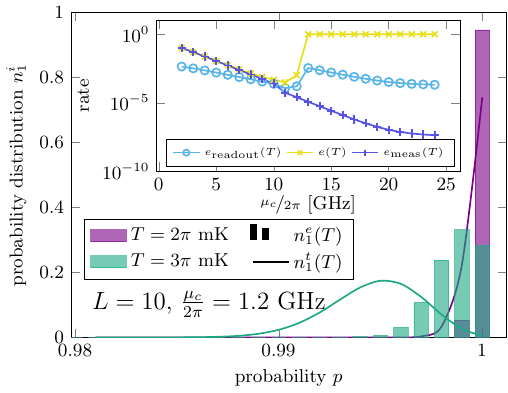}%
			}%
		}%
		\hfill%
		\subfloat[\label{fig:fidelityL20}]%
		{%
			\ifthenelse{\boolean{buildtikzpics}}%
			{%
				\def\Lvalue{20}%
				\def\muValue{17}	
				\input{figures/fidelity}%
			}%
			{%
				\includegraphics{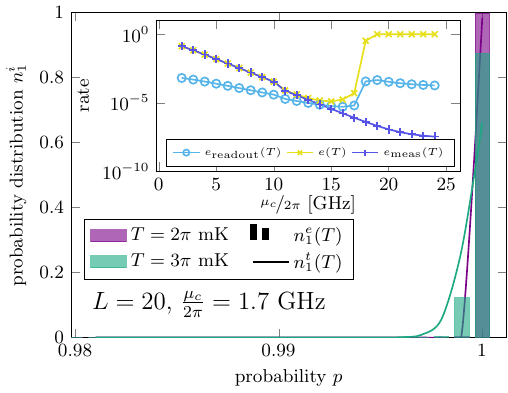}%
			}%
		}%
		\caption%
		{%
			\label{fig:fidelity}%
			Comparison of theoretical and experimental probability distribution for model parameters $s/2\pi = 400\, \mathrm{MHz}$, $s^\prime/2\pi = 1\, \mathrm{MHz}$.
			We show data for $L=10 (20)$ stabilizer qubits with detunings $\mu_c/2\pi = 1.2 (1.7)\, \mathrm{GHz}$ in \cref{fig:fidelityL10} (\cref{fig:fidelityL20}) at half filling for $T = 2\pi,3\pi\,\mathrm{mK}$.
			The bar plots show the experimental probability distribution to obtain an occupied control qubit from the simulated measurement of the control qubit, $n_1^e$, in a sequence of $M = 10^3$ independent excitation processes and averaged over an ensemble of $K = 10^3$ independent realizations.
			The line plots correspond to the theoretical probability distribution $n_1^t(T) = B(M,p = F(T))$.
			The inset shows the theoretical error rate $e(T)$ (yellow curve) as well as the experimental error rate $e_{\mathrm{meas}}(T)$ (dark blue curve) as a function of the on site potential of the control qubit $\mu_c$.
			The geometric mean of both curves, $e_{\mathrm{readout}}$ (light blue curve), suffices as a calibration function.
			Minimizing this curve leads to the optimal value of the detuning $\mu_c$.
		}%
	\end{figure}
	In the previous analysis it was assumed that the fidelity of an excitation of the logical qubit can be identified by determining the weight of the $\ket{1}$ states in the post\hyp excitation density operator.
	However, in practise one measures the occupation of the control qubit after an excitation (see~\cref{fig:many_body_spectrum_occupancies}).
	We identify states in the lower branch of the $N_{k_0} = 2$ sector via an occupied control qubit, and thereby verify if an excitation on the control qubit has successfully created the excited state $\ket{1}$ of the logical qubit.
	However, there can be false positives, i.e., the control qubit can be occupied although the corresponding eigenstate is not in the desired lower branch of the $N_{k_0}=2$ sector.
	To deduce the impact of these false positives, we compare the theoretically obtained probability distribution to find a state in the desired low\hyp lying $N_{k_0}=2$ sector to the probability distribution to measure an occupied control qubit.
	The theoretical distribution is given by a binomial distribution characterized by the fidelity $n_1^t(T) = B(M,p = F(T))$, with $F(T)$ defined in \cref{eq:wheel+probe:fideliT} and $M$ being the number of trials.
	To obtain the experimentally accessible probability distribution, we simulate a sequence of $M$ independent measurement processes on the thermal density operator $\hat \rho_1(T)$ at different temperatures $T$, yielding $N_1(T) \leq M$ cases of occupied control qubits.
	Averaging $N_1(T)$ over an ensemble of $K$ independent realizations provides access to the experimentally observed distribution of the relative incidences $n^e_1(T) = N_1(T)/M$.
	\cref{fig:fidelity} compares the experimental and theoretical distribution for temperatures $T = 2\pi,3\pi\,\mathrm{mK}$ and $L = 10 (20)$ stabilizer qubits at detunings $\mu_c/2\pi = 1.2 (1.7)\, \mathrm{GHz}$.
	For $L = 10$,  see~\cref{fig:fidelityL10}, the impact of false positives can be observed clearly as a shift of the mean towards higher incidences, compared to the theoretical distribution.
	For the smallest temperature considered ($T=2\pi\,\mathrm{mK}$), the experimentally observable fidelity overestimates the actual fidelity to excite into the desired $N_{k_0}=2$ branch by approximately $3$ out of $1000$ samples.
	Increasing $L$ to $L=20$ allows to obtain better agreement of both distributions, see~\cref{fig:fidelityL20}.
	Again, this can be explained by geometric stabilization, which allows to suppress the effect of false positives caused by states in other clusters.
	To further increase the agreement, it is possible to choose $\mu_c$ in an optimal way by minimizing the calibration function
	\begin{align}
		e_{\mathrm{readout}}(T) = \sqrt{\abs{e(T) - e_{\mathrm{meas}}(T)}\cdot e(T)}\;.
	\end{align}
	Here, $e(T) = 1-F(T)$ denotes the theoretical error rate and $e_{\mathrm{meas}}(T)$ the experimental error rate, i.e., the probability of the control qubit not being occupied in a measurement.
	We defined the calibration function as the geometric mean of both rates.
	As a key result, we observe that for the optimal detuning $\mu_c/2\pi \approx 1.7\, \mathrm{GHz}$, error rates $e< 10^{-3}$ can be obtained, which is shown in the insets of \cref{fig:fidelity}.
	Note that in these computations we neglected the impact of noisy wheel\hyp to\hyp center couplings, which is justified by the results in~\cref{sec:exp-proposal}, showing that for practical realizations the modifications of the many\hyp body spectrum are irrelevant compared to the energy shifts introduced from geometrical stabilization.
	\subsection{\label{sec:logical-qubit:non-local} Non\hyp local coupling protocol}
	We now turn to the non\hyp local protocol realizing the qubit\hyp subspace via a three\hyp body interaction.
	For that purpose, we introduce the operator
	\begin{equation}
		\hat K_{k_0} = \sum_{j=0}^{L-1} \mathrm e^{-\mathrm i k_0 j}\hat h^\nodagger_j \;.
	\end{equation}
	Note that $\hat K^{(\dagger)}_{k_0}$ removes (adds) a collective excitation with momentum $k_0$ to the set of stabilizer qubits.
	This operator forms the building block for non\hyp local operations moving excitations from the control qubit, via the cavity into the stabilizer qubits (see also~\cref{fig:wheel_nonlocal}).
	Specifically, we consider the following extension to~\cref{eq:hubb-wheel}:
	\begin{equation}
		\hat H_\mathrm{qubit}
		=
		\hat H_\mathrm{wheel} + s^\prime \left(\hat K^\dagger_{k_0} \hat h^\nodagger_\odot \hat h^\nodagger_c + \mathrm{h.c.} \right) + \mu_c \hat n_c \;. \label{eq:wheel+probe:hamiltonian:non-local}
	\end{equation}
	For now, we set the chemical potential to $\mu_c = 0$.
	We note that 
	\begin{equation}
		[\hat K^\dagger_{k_0}\hat h^\nodagger_\odot, \hat n_{k_0}] = 0\;,
		\label{eq:nonlocal_commutator}
	\end{equation}
	and thus $\hat K^\dagger \hat h^\nodagger_\odot$ leaves the occupation of the bright modes invariant.
	It does, however, couple the eigenstates $\ket{\pm}$ of the~\gls{BHW} in the bright subspace.
	\begin{figure}
		\centering
		\ifthenelse{\boolean{buildtikzpics}}%
		{%
			\input{figures/spectrum_non_local_coupling}
	  	}
  		{
  			\includegraphics{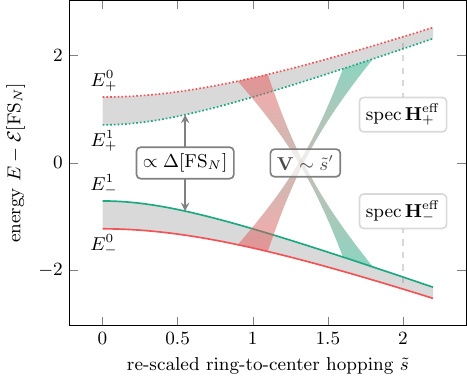}%
  		}
		\caption{
			\label{fig:spectrum:non-local-coupling}
			Illustration of a non-trivial four\hyp dimensional sub block $\mathbf H^\mathrm{eff}_Q$ from the non\hyp local coupling protocol for a specific sector, fixed by a Slater determinant $\mathrm{FS}_N$ (c.f.~\cref{eq:wheel+probe:non-local:sub-block}).
			Spectral ranges of the effective $2\times 2$ Hamiltonians $\mathbf H^\mathrm{eff}_\pm$ are indicated by the gray shaded regions.
			Red/Green lines indicate the marginal energies $E^{0/1}_\pm$ confining the two\hyp dimensional manifolds of each effective Hamiltonian.
			The red/green shaded ribbons mark the hybridization $\mathbf V$ between the manifolds.
			Note that for illustrative purposes we set $\tilde s^\prime = 1$ in this plot, because the spectral ranges $\operatorname{spec} \mathbf H^\mathrm{eff}_\pm \sim \tilde s^\prime$ are too narrow in the practically relevant case $\tilde s^\prime \ll \Delta[\mathrm{FS}_N]$.
		}
	\end{figure}
	We expand~\cref{eq:wheel+probe:hamiltonian:non-local} in the tensor product space spanned by the states~\cref{eq:hubb-wheel:estate:0,eq:hubb-wheel:estate:pm,eq:hubb-wheel:estate:2} and the control qubit states $\ket{0_c}, \ket{1_c}$.
	Crucially, due to~\cref{eq:nonlocal_commutator} these basis states are defined w.r.t. a fixed Slater determinant $\ket{\mathrm{FS}_N}$, in contrast to its analogon in the local coupling protocol~\cref{eq:wheel:probe:mbstates}.
	In this basis, the Hamiltonian is block\hyp diagonal for each Slater determinant and the only non\hyp trivial block couples the states $\ket{0_c,+}, \ket{1_c,+}, \ket{0_c,-}, \ket{1_c,-}$, which reads
	\begin{align}
		\mathbf H_Q^\mathrm{eff} - \mathcal E[\mathrm{FS}_N] \mathbb 1_{4\times 4}
		&=
		\begin{pmatrix}
			\mathbf H^\mathrm{eff}_+ & \mathbf V \\ \mathbf V^\dagger & \mathbf H^\mathrm{eff}_- \; .
		\end{pmatrix}
		\label{eq:wheel+probe:non-local:sub-block}
	\end{align}
	Here, $\mathcal E[\mathrm{FS}_N] = E[\mathrm{FS}_N](1-1/L)+1$ and $E[\mathrm{FS}_N]=\sum_{k\in \mathrm{FS}_N}\varepsilon_k$ is the energy contribution from the $N$ single\hyp particle states contained in the spectator Slater determinant $\mathrm{FS}_N$.
	The exact definition of the matrix $\mathbf V$ is given in~\cref{app:nonlocal-protocol}.
	For the subsequent discussion, the only relevant property is that it controls the hybridization between the two manifolds $\ket{0/1_c,+}$ and $\ket{0/1_c,-}$ in the bright subspace, as well as its parametic dependency $\lVert \mathbf V \rVert \sim \tilde s^\prime$ on the energy scale $\tilde s^\prime = \sqrt{L} s^\prime$.
	In~\cref{fig:spectrum:non-local-coupling} we illustrate the structure of~\cref{eq:wheel+probe:non-local:sub-block}.
	For a given Slater determinant $\mathrm{FS}_N$, the non\hyp local protocol confines the dynamics into independent, four\hyp dimensional sectors described by $\mathbf H^\mathrm{eff}_Q$.
	Within each sector, the bright states $\ket{0_c/1_c,+}$ and $\ket{0_c/1_c,-}$ form two manifolds with Hamiltonians $\mathbf H^\mathrm{eff}_\pm$ separated by a many\hyp body gap
	\begin{equation}
		2\Delta[\mathrm{FS}_N]
		=
		2 \sqrt{(E[\mathrm{FS}_N]/L + 1)^2 + \tilde s^2} \;.
	\end{equation}
	The effect of $\mathbf V$ is then to couple states with different occupations of the control qubit between the manifolds, i.e., it hybridizes $\ket{0_c,+} \leftrightarrow \ket{1_c,-}$ and $\ket{1_c,+} \leftrightarrow \ket{0_c,-}$.
	In the regime $\tilde s^\prime \ll \Delta[\mathrm{FS}_N] \sim \tilde s$, virtual transitions between the two manifolds can be eliminated perturbatively.
	Let us define the transformed hybridization $\mathbf W = \mathbf X^\dagger_+ \mathbf V \mathbf X_-$, as well as $\mathbf \Lambda_\pm = \operatorname{diag}(E^0_\pm, E^1_\pm)$, where $E^{0/1}_\pm$ are the eigenvalues of $\mathbf H^\mathrm{eff}_\pm$.
	Then, to order $\mathbf V^2$ the decoupled manifolds are described by
	\begin{align}
		\mathbf H^Q_+ &= \mathbf \Lambda_+ + \frac{1}{2}\left(\mathbf W \mathbf S^\dagger + \mathbf S \mathbf W^\dagger\right)\;,\\
		\mathbf H^Q_- &= \mathbf \Lambda_- - \frac{1}{2}\left(\mathbf W^\dagger \mathbf S + \mathbf S^\dagger \mathbf W\right)\;,
	\end{align}
	where $S_{nm} = \frac{W_{nm}}{E^n_+ - E^m_-}$ is the generator of a Schrieffer-Wolf transformation.
	The resulting lower manifold constitutes an effective two\hyp level system whose Hamiltonian can be written as
	\begin{equation}
		\mathbf H^Q_-
		=
		\begin{pmatrix}
			\varepsilon^0 & J_\mathrm{eff} \\
			\overbar{J}_\mathrm{eff} & \varepsilon^1
		\end{pmatrix}
		\equiv
		\overbar{\varepsilon} \mathbb 1_{2\times 2} + 
		\begin{pmatrix}
			\delta & J_\mathrm{eff} \\
			\overbar{J}_\mathrm{eff} & -\delta
		\end{pmatrix} \;,
		\label{eq:wheel+probe:non-local:logical-qubit}
	\end{equation}
	where we introduced the mean energy $\overbar{\varepsilon} = (\varepsilon^0 + \varepsilon^1) / 2$ as well as the splitting $\delta = (\varepsilon^0 - \varepsilon^1) / 2$.
	The separation from the eliminated manifold provides energetic protection against leakage with corrections $\sim \mathcal{O}(\lVert \mathbf V \rVert^3/\Delta^2[\mathrm{FS}_N])$, while the parameters $\delta$ and $J_\mathrm{eff}$ describe, respectively, differential energy shifts and coherent rotations within the logical subspace.
	We refer to~\cref{app:nonlocal-protocol} for a detailed derivation.
	The couplings in $\mathbf H^Q_-$ depend on the external parameters, i.e., $J_\mathrm{eff} \equiv J_\mathrm{eff}(t/\tilde s,\mu_c)$ and $\delta \equiv \delta(t/\tilde s,\mu_c)$.
	A particularly simple operating point is obtained for $t=\mu_c \equiv 0$, which gives a highly symmetric inter-manifold coupling.
	In this limit, we find $\mathbf W = -\frac{\mathrm i\tilde s^\prime}{2}\mathbf \sigma_y$ and therefrom
	\begin{equation}
		\mathbf H^Q_-
		\simeq
		\overbar{\varepsilon} \mathbb 1_{2\times 2} - \frac{\mathbf W^\dagger \mathbf W}{2\Delta[\underbar{k}_N]}
		=
		\left(\overbar{\varepsilon} - \frac{(\tilde s^\prime)^2}{8\Delta[\mathrm{FS}_N]}\right)\mathbb 1_{2\times 2}
	\end{equation}
	up to order $\sim \mathcal O((\tilde s^\prime)^3/\tilde s^2$.
	Virtual exchanges with the high\hyp energy manifold therefore produce, to leading nonvanishing order, only a common energy renormalization of the two low\hyp energy states.
	%
	%

	%
	Next, let us consider the effect of noise altering one of the external parameter $p\in \{t/\tilde s, \mu_c\}$ at a~\gls{SOP}.
	Upon introducing a perturbation $p\rightarrow p+\Delta p$, the effective qubit parameters are modified according to $\delta\rightarrow \delta + \Delta \delta(\Delta p)$ and $J_\mathrm{eff} \rightarrow J_\mathrm{eff} + \Delta J_\mathrm{eff}(\Delta p)$.
	Thus, to first order the induced perturbations are
	\begin{equation}
		\Delta \delta = \left.\frac{\partial \delta}{\partial p}\right\vert_{\Delta p=0} \Delta p\;,
		\quad
		\Delta J_\mathrm{eff} = \left.\frac{\partial J_\mathrm{eff}}{\partial p}\right\vert_{\Delta p=0}\Delta p \;.
	\end{equation}
	Whenever at least one of these derivatives vanishes, we call the pair $(t,\mu_c)$ a~\gls{SOP}.
	The important consequence is that at a~\gls{SOP}, phase errors are suppressed to leading order $\sim \mathcal O(\Delta p\cdot \sqrt{L} (s^\prime)^3/s^2)$ in the external perturbation~\cite{Koch2007,Fei2015,Huang2021}.
	\begin{figure}
		\centering
		\ifthenelse{\boolean{buildtikzpics}}%
			{%
				\input{figures/effective_model_couplings}
			}%
			{%
				\includegraphics{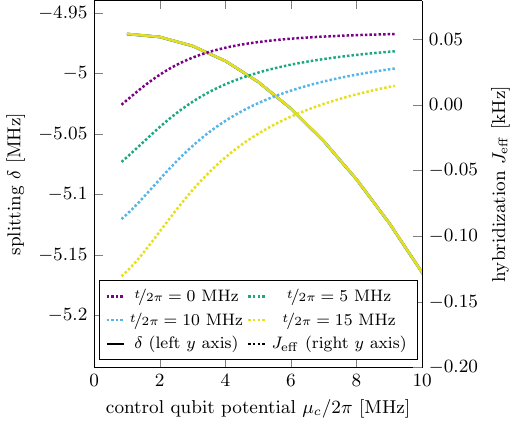}%
			}%
		\caption{
			\label{fig:effective-non-model:couplings}
			Effective qubit Hamiltonian couplings evaluated at $s^\prime/2\pi = 1\mathrm{MHz}$ and $s/2\pi = 100\mathrm{MHz}$, as a function of the external parameters $t,\mu_c$.
			Left $y$\hyp axis shows the qubit detuning in MHz (solid lines), right $y$\hyp axis the qubit hybridizaion in kHz (dotted lines).
			Note the vanishing derivative $\partial \delta / \partial \mu_c$ near $\mu_c = 0$.
		}
	\end{figure}
	The same microscopic parameters that determine the~\gls{SOP} can subsequently be used for qubit control.
	A finite chemical potential $\mu_c$ produces a controllable splitting $\delta$ between the logical states.
	This allows to implement phase gates by a simple time\hyp evolution.
	Changing the hopping $t$ of the~\gls{BHW} modifies the finite effective tunnelling amplitude $J_\mathrm{eff}$.
	An evolution of the whole system for a time $T_X = \pi/2n\vert J_\mathrm{eff}\vert$ then produces an $X$ rotation with angle $\pi/n$.
	However, changing either the hopping or the chemical potential generally modifies both $J_\mathrm{eff}$ and $\delta$.
	As a consequence, high-fidelity gate operations, which are implemented by tuning one effective qubit parameter, always require specific choices of both external parameters to compensate the unwanted induced perturbations in the respective other effective qubit parameter.
	In~\cref{fig:effective-non-model:couplings} we demonstrate these dependencies for a fixed probe\hyp site coupling $s^\prime/2\pi = 1\mathrm{MHz}$.
	Note that the hybridization $J_\mathrm{eff}$ is indeed vanishing at the~\gls{SOP} $\mu_c=0$.
	We highlight two important conclusions that can be drawn from the data.
	First, the hybridizations and the detunings satisfy $J_\mathrm{eff}/\delta \sim s^\prime$, which is a direct consequence of them being effective couplings derived from the Schrieffer\hyp Wolff transformation.
	This implies that the qubit switching times are $\sim 10\mathrm{ms}$ in this example.
	Taking into account that dependence of the splitting is practically independent on the hopping parameter, the nonlocal protocol provides high\hyp precision and independent microscopic control over the two effective Bloch\hyp sphere axes.
	A second observation is that the derivative of the detuning $\delta$ vanishes: $\partial \delta/\partial \mu_c \rightarrow 0$ for $\mu_c,J_\mathrm{eff}\rightarrow 0$.
	This indeed shows that $\mu_c=0$ provides a~\gls{SOP} w.r.t. the logical qubit splitting $\delta$.
	As a consequence, disorder effects on the detuning induced by perturbations in the chemical potential of the control qubit only contribute at second order in $\sqrt L (s^\prime)^3/s^2$.
	\subsection{Stability of coherent superpositions}
	The two coupling protocols constitute different approaches to protect coherent superpositions within the logical qubit subspace.
	Both schemes fundamentally rely on the structure of the many\hyp body spectrum, i.e., the many\hyp body gap induced by the separation of the bright and dark states.
	This effect creates an energetic protection of the low\hyp lying logical qubit manifold $~\sim\mathcal O(\tilde s)$ and we discussed the corresponding protection against perturbations incurred by finite temperatures in~\cref{sec:logical-qubit:local}.
	However, there are also significant differences between both protocols, which we discuss in the following.
	The local coupling protocol is based solely on the fact that the logical information itself is represented via the occupation of the control qubit.
	Thus, generating superpositions by applying rotations on the control qubit, the classical probabilities of the Slater determinants in the mixed state remain unaffected.
	Furthermore, even though the Slater determinants are energetically close, or even degenerate in the limit $t/s\rightarrow 0$, their unitary dynamics can not leave the logical\hyp qubit branches without coupling to the trivial branch, which is energetically separated by the many\hyp body gap $\sim \tilde s$.
	As a consequence, we may conclude that only those perturbations are relevant, which either reduce the many\hyp body gap, i.e., excitations on the stabilizer qubits, or excitations acting directly on the control qubit.
	The latter type of error is energetically suppressed by the chemical potential $\mu_c$ and, thus, can be neglected if we assume values $\mu_c/2\pi \sim 1\,\operatorname{GHz}$ corresponding to a temperature of $100\pi\,\operatorname{mK}$.
	The energetically more relevant sources are thus those, which reduce the many\hyp body gap by reducing the number of particles in the Slater determinants $N-N_{k_0}$, i.e., the occupation of the stabilizer qubits.
	In order to investigate the decay of this occupation, it is possible to simulate the dynamics of an excited logical qubit in the presence of dissipation acting on the stabilizer qubits.
	Future work will be devoted to a further discussion.
	However, the local protocol does not provide protection against dephasing errors.
	The non\hyp local coupling protocol follows a different strategy.
	Here, the logical subspace is formed by the highly non\hyp local eigenstates of~\cref{eq:wheel+probe:non-local:logical-qubit}.
	Crucially, this logical qubit Hamiltonian can be controlled via the external parameters $t/\tilde s,\mu_c$.
	In the regime of small couplings between the center site and the control qubit $s/s^\prime$, our calculations showed that the two parameters provide tuning knobs to control the effective qubit parameters independently.
	Moreover, there are~\glspl{SOP} such that first order fluctuations in the external parameters can be cancelled and, thereby, dephasing of coherent superpositions is suppressed up to second order in $\sqrt L (s^\prime)^3/s^2$.
	Therefore, the non\hyp local protocol exploits the many-body gap and a symmetry of the effective hybridization to suppress first-order differential energy shifts.
	\section{\label{sec:conclusion}Conclusion and Outlook}%
	We proposed a logical qubit construction scheme based on the peculiar property of the~\gls{BHW} to open an extensive gap between clusters of many\hyp body eigenstates with the energetically lowest cluster being composed of superradiant bright states~\cite{Dicke1954,Wilke23}.
	The energetic separation $\sim \tilde s = s\sqrt L$ of this low\hyp lying cluster is controlled by an all\hyp to\hyp one coupling of magnitude $s$ between stabilizer qubits and a central two\hyp level system.
	We note that such an all\hyp to\hyp one\hyp connectivity between \glspl{SCQ} and a two-level resonator has been realized only recently~\cite{Renger2026}.
	This energetically protected cluster forms the backbone of a quantum memory, and we show that gaussian noise acting the stabilizer qubits perturbs the many\hyp body spectrum only at second order in the standard deviation $\sigma$.
	In order to utilise this memory as a logical qubit, we introduce a control qubit to create two logical subspaces.
	Crucially, these two subspaces remain energetically well separated from the remaining many\hyp body spectrum via the~\gls{BHW} many\hyp body gap.
	We then investigate two different coupling schemes for the control qubit.
	A possible implementation of the first scheme can leverage on high\hyp connectivity~\gls{SCQ} architectures such as the star geometry, in which~\glspl{SCQ}are logically coupled to a central cavity mode~\cite{Renger2026}.
	In that case, the logical qubit subspace is formed via a local coupling $s^\prime$ between the control qubit and the central two\hyp level system, and the introduction of a chemical potential $\mu_c$ on the control qubit, which can be implemented via a frequency detuning w.r.t. the stabilizer qubits~\cref{eq:wheel+probe:hamiltonian:local}.
	In the weak\hyp coupling limit $s^\prime \ll s$, we show that the resulting logical qubit can be read out and switched via local operations on the control qubit, only.
	For a single\hyp qubit $X$\hyp gate acting on the logical ground state, we computed the fidelity $F(T)$ and showed that for experimentally feasible transmon frequencies, temperatures $T \sim 3\pi \approx 10\,\mathrm{mK}$ are sufficient to reach theoretical error rates $e(T) = 10^{-4}$, using $L=20$ stabilizer qubits.
	In these calculations, the relevant quantity is the ratio between the renormalized coupling $\Tilde s$ of the stabilizer\hyp qubits to the central mode and the frequency detuning between stabilizer qubits and control qubit $\mu_c$.
	This can be exploited to adapt to practical constraints, for instance a reduction of the central mode's frequency when increasing its length in order to increase the number of stabilizer qubits.
	We also analyzed the occurrence of read\hyp out errors (false positives) and introduced a calibration function, which can be used to experimentally vary the detuning such that the combined false\hyp positive\hyp{} and theoretical error\hyp rate is minimized.
	Given a certain number of stabilizer qubits, the calibration function therefore allows to tune the logical qubit such that for the discussed parameters ($L=20$, $T\approx 10\,\mathrm{mK}$), it can be operated with minimal error rates $<10^{-4}$ using a frequency detuning of the control\hyp site qubit of $\mu_c \approx 1.7\,\mathrm{GHz}$.
	The second scheme is designed to further suppress dephasing noise within the logical qubit manifold.
	For that purpose, we investigate a non\hyp local coupling $s^\prime$ between the control qubit, the central mode and a collective excitation of the stabilizer qubits~\cref{eq:wheel+probe:hamiltonian:non-local}.
	The extensively scaling many\hyp body gap of the~\gls{BHW} allows for a perturbative treatment of the resulting problem, which again separates into clusters of many\hyp body eigenstates separated by the~\gls{BHW} many\hyp body gap $\sim \tilde s$.
	In the low\hyp energy subspace, an effective two\hyp level system is obtained whose couplings are determined by the external parameters $t,\mu_c$.
	We derived the existence of~\gls{SOP} at which dephasing noise is suppressed to second order in $\sim \sqrt L (s^\prime)^3/s^2$.
	Moreover, the logical qubit can be controlled on the Bloch sphere utilizing the dependency of the effective hybridization between the logical qubit states on the external hopping parameter $t$.
	We demonstrate that for a suitable choice of parameters $s/2\pi=100\mathrm{MHz},t/2\pi=15\mathrm{MHz},s^\prime/2\pi=1\mathrm{MHz}$, the timescale to switch the logical qubit state is of the order of $\sim 10\mathrm{ms}$, yet faster switching times can be achieved by controlling the ratio $s^\prime/s$.
	Our analysis of the robustness of the two coupling schemes already suggests a high degree of control to suppress the effects of perturbations introduced by experimental imperfections and temperature.
	Nevertheless, our considerations are based on two critical assumptions: (i) the constituting qubits are ideal two\hyp level systems and (ii) there is only one excitation in the central mode at the most.
	For weak violations of the first assumption, i.e., a small subset of the stabilizer qubits forming the wheel exhibit a loss of coherence or excitations into energetically higher states, we still expect our results to be valid.
	This is based on the fact that the logical qubit\hyp state is encoded with a high redundancy in a cluster of many\hyp body eigenstates whose number scales exponentially with the wheel's coordination number.
	The effects of the anharmonic transmon spectrum could be studied numerically using open quantum\hyp system methods:
	A likewise interesting and important future task includes simulating the logical qubit dynamics resulting from decay of the individual physical qubits, which can be performed by simulating dissipation on the ring sites of the \gls{BHW}.
	The second assumption could be realized to a very high degree via fine\hyp tuning the stabilizer qubits to resonance with the central mode and choose a large qubit frequency, compared to the coupling strength $s$.
	Alternatively, state\hyp of\hyp the\hyp art digital quantum simulators allow to enforce an upper bound on the central mode occupation~\cite{Renger2026}.
	Furthermore, as a practical source of imperfections the quality factor of the central mode would have to be considered.
	Despite these open questions, this work demonstrates that the specific characteristics of the~\gls{BHW} open up new perspective to the design of error\hyp resilient logical qubit architectures.
	Geometric stabilization provides energetic protection, while the structure of the coupling protocol determines whether additional protection against logical errors can be achieved.
	The robust clustering of many\hyp body eigenstates provides the practical means to engineer large energy gaps, i.e., geometric protection of the low\hyp energy qubit subspace.
	Based on this property, we show that within that subspace logical qubits can be designed, which, depending on the experimental platform, exhibit additional error resilience.
	While the local coupling protocol can be readily implemented~\cite{Renger2026}, the non\hyp local coupling protocol is more challenging.
	However, neutral atom arrays or trapped ion systems could provide the necessary non\hyp local coupling schemes.
	For these platforms, our model opens a possibility to achieve passive error protection against both, energetic and parametric noise.
	\section*{\label{sec:acknowledgement}Acknowledgment}%
	We are indebted to Philip Kim for insightful discussions and attentively reading the manuscript. %
	Furthermore, we want to express our gratitude to Frank Deppe and Miha Papi\u c for both, sharing their practical insights and fruitful discussions.
	We thank Alexander Rommens, Felix Palm and Henning Schl\"omer for carefully reading the manuscript. %
	We acknowledge Lena Scheuchl for improving our understanding of the~\gls{BHW} with control qubit. %
	RHW and SP acknowledge support from the Munich Center for Quantum Science and Technology. %
	TK acknowledges financial support by the ERC Starting Grant from the European Union’s Horizon 2020 research and innovation programme under grant agreement No 758935; and the UK’s Engineering and Physical Sciences Research Council [EPSRC; grant number EP/W022982/1].
	The computations were enabled by resources in project NAISS 2023/22-527 provided by the National Academic Infrastructure for Supercomputing in Sweden (NAISS) at UPPMAX, funded by the Swedish Research Council through grant agreement no. 2022-06725.
	\bibliography{Literatur}%
	\clearpage
	\widetext
\begin{center}
\textbf{\large Supplemental Materials}
\end{center}
\setcounter{equation}{0}%
\setcounter{figure}{0}%
\setcounter{table}{0}%
\setcounter{page}{1}%
\makeatletter%
\renewcommand{\theequation}{S\arabic{equation}}%
\renewcommand{\thefigure}{S\arabic{figure}}%
\renewcommand{\bibnumfmt}[1]{[S#1]}%
\makeatother
%
%

\section{Solution Strategy of the Bose\hyp Hubbard wheel\label{app:solution_BHW}}
\begin{figure}
	\centering
	\ifthenelse{\boolean{buildtikzpics}}%
	{%
		\input{figures/subspaces}%
	}%
	{%
		\includegraphics{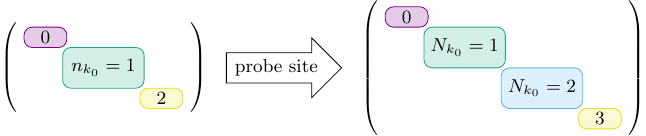}%
	}%
	\caption{Blockdiagonal structure of the Hamiltonian in $n_{k_0}$ (left, \cref{eq:sm:hubb-wheel}) before and in $N_{k_0}$ (right, \cref{eq:sm:hubb-wheel-probe}) after adding the control qubit to the system. 
		The matrix on the right is spanned by all possible combinations of states with occupations $n_{k_0} = 0,1_\pm,2 $ and $n_c = 0,1$.
		Hence, the blocks have the size $1-2-1$ and $1-3-3-1$.
		To provide an example, in the $N_{k_0}=0,3$ subspaces the corresponding states have the form $\ket{n_c=0}\ket{n_{k_0}=0}, \ket{n_c=1}\ket{n_{k_0}=2}$, respectively.}
	\label{fig:sm:subspaces}
\end{figure}

The Hamiltonian of the Bose\hyp Hubbard wheel is given by
\begin{align}%
	\hat H_{\mathrm{wheel}} = -t \sum_{j=0}^{L-1} \left( \hat h^\dagger_j \hat h^\nodagger_{j+1} + \mathrm{h.c.} \right) -%
	\sum_{j=0}^{L-1} \left( s_j\hat h^\dagger_j \hat h^\nodagger_\odot + \mathrm{h.c.} \right)\;,
	\label{eq:sm:hubb-wheel}%
\end{align}%
where $\hat h^{(\dagger)}_j$ denote the hardcore bosonic annihilation (creation) operators on site $j$ on the ring of the wheel while $\hat h^{(\dagger)}_\odot$ corresponds to the annihilation (creation) of \gls{HCB} on the center site.
$L$ denotes the number of sites on the ring and we consider periodic boundary conditions.
Direct application of common solution strategies fail due to long\hyp ranged hopping introduced by the center cite when mapping the system to a chain. 
Our solution strategy is based on a geometric ansatz that maps the Bose\hyp Hubbard wheel to a spinless fermion ladder with periodic boundary conditions.
On this extended Hilbert space, the full many\hyp body problem can be solved and eventually projected down to the initial Hilbert space.
Please note that a more detailed derivation of the solution strategy can be found in our recent publication \cite{Wilke23}.
The resulting many\hyp body eigenstates have the structure
\begin{align}
	\ket{n_{k_0}}\ket{FS_{N-n_{k_0}}}\;,
\end{align}
where $n_{k_0} = 0, 1_\pm, 2$ denotes the occupation of a distinct mode and the entire many\hyp body problem reduces to the diagonalization of a $4\times 4$ matrix, see left side of \cref{fig:sm:subspaces}.
The occupation of the $k_0$ mode has significant effects on both the single\hyp and many\hyp particle spectrum, as it gaps out $\propto s\sqrt{L}$ while all remaining modes follow a common tight\hyp binding dispersion.
Furthermore $n_{k_0}$ generates a $\mathbb{Z}_2$ symmetry in the many\hyp body spectrum distinguishing \gls{BEC} states, which gap out $\propto s\sqrt{L}$ as a result of this characteristic, and non\hyp \gls{BEC} states.
$\ket{FS_{N-n_{k_0}}}$ denotes a Slater determinant of $N-n_{k_0}$ $k\neq k_0$ modes with $k \in \left\{ k_n=\frac{2\pi}{L}n | n=0,\ldots, L-1 \right\}$.
The eigenstates of the projected wheel are given by
\begin{align}
	&\ket{E_{n_{k_0} = 0}(\mathrm{FS}_{N})} = \ket{0}\ket{FS_{N}}\;,\\
	&\ket{E_{n_{k_0} = \pm}(\mathrm{FS}_{N-1})} = \ket{\pm}\ket{FS_{N-1}}\;,\\
	&\ket{E_{n_{k_0} = 2}(\mathrm{FS}_{N-2})} = \ket{2}\ket{FS_{N-2}}\;,
\end{align}
where $\ket{\pm}$ diagonalize the $n_{k_0} = 1$ subspace.
These expressions, as well as the analytically obtained many\hyp particle energies for the $4\times 4$ $n_{k_0}$ subspace
\begin{align}
	E_0(FS_{N}) &= E(FS_N), \\
	E_{ \pm} (FS_{N-1},k^\prime) &= E(FS_{N-1}) \left(1-\frac{1}{L} \right) + 1 \pm \sqrt{\left(\frac{E(FS_{N-1})}{L} + 1\right)^2 + \tilde s^2}, \\
	E_{2} (FS_{N-2},k^\prime, k^\pprime) &= \left(E(FS_{N-2}) +2 \right) \left(1-\frac{2}{L} \right) \; ,
	\label{eq:sm:hubb-wheel:mp-eigenvalues}
\end{align}
will be needed for the solution strategy of the wheel\hyp probe system.
Here, $k^\prime, k^{\prime\prime}$ denote the $k_0$ modes.
Note that these energies depend on the choice of Slater determinant $FS_N$ outside of the $n_{k_0}$ subspace.\\
For later convenience, we state the eigenstates $\ket{\pm}$ for a given Slater determinant
\begin{align}
	\ket{\pm}   = c^\pm_{1+}\ket{1_+} + c^\pm_{1-}\ket{1_-}  = d^\pm_0 \ket{n_{k_0,O} = 1}\ket{n_\odot = 0} + d^\pm_1 \ket{n_{k_0,O} = 0}\ket{n_\odot = 1}\;,
	\label{eq:sm:pm_coeffs}
\end{align}
where $d^\pm_0 = c^\pm_{1+} \psi_+ + c^\pm_{1-} \psi_-$ and $d^\pm_1 = c^\pm_{1+} \psi_+ \Delta_+ + c^\pm_{1-} \psi_-\Delta_-$.
Here, $n_{k_0,O} (n_\odot)$ denotes the occupation of the $k_0$ mode on the outer ring (center site) of the wheel.
The coefficients are given by $\psi_\pm = \frac{1}{\sqrt{1+\abs{\Delta_\pm}^2}}$, $\Delta_\pm = \frac{\epsilon_0}{2\tilde{s}} \pm \frac{\sqrt{\epsilon^2_0 + 4\tilde{s}^2}}{2\abs{\tilde{s}}}$ and $\epsilon_0 = 2t\cos{k_0}$.
The remaining states take the form
\begin{align}
	\ket{0} =\ket{n_{k_0,O} = 0}\ket{n_\odot = 0} , \ket{2} = \ket{n_{k_0,O} = 1}\ket{n_\odot = 1}\;.
\end{align}

\section{\label{app:noisy-wheel}Noise}

\begin{figure*}[t]%
	\centering%
	\ifthenelse{\boolean{buildtikzpics}}%
	{%
		\input{figures/noisy_single_particle_spectrum_self_consitency}%
	}%
	{%
		\includegraphics{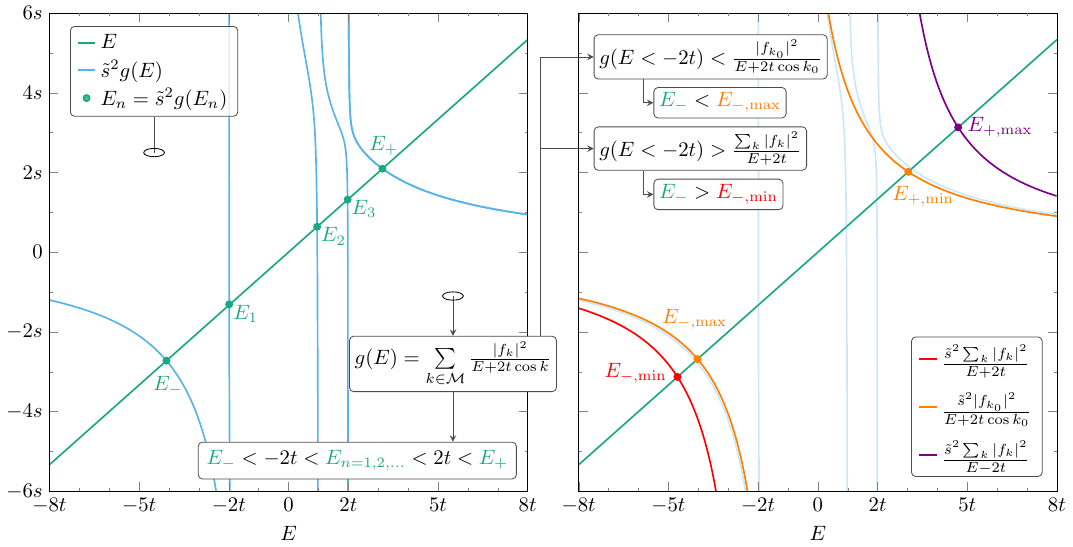}%
	}%
	\caption
	{%
		\label{fig:wheel:disorder:sp:g}%
		Graphical representation of Eq.~\eqref{eq:wheel:disorder:sp:g} for a system with $L = 6$ lattice sites on the outer ring, $k_0 = \pi / 3$, $s/t = 1.5$ and random coefficients $s_j$ with $\sigma/t = 0.4$.
		The solutions $E_n$ are given by the intersections of the blue and green graph in the left panel.
		The best bounds for $E_-$ are obtained from evaluating~\cref{eq:wheel:disorder:sp:e_minus_upper} at $k=k_0$ and solving for the intersections, which are marked by $E_\mathrm{-,min}$ and $E_\mathrm{-,max}$ in the right panel.
		Similarly, the corresponding best estimations for the bounds on $E_+$ are marked by $E_\mathrm{+,min}$ and $E_\mathrm{+,max}$.
	}%
\end{figure*}%
Here, we outline the analysis of the Bose\hyp Hubbard wheel's spectrum in the presence of noise in detail.
Mapping the hardcore\hyp bosonic operators $\hat h^{(\dagger)}_j, \hat h^{(\dagger)}_\odot$ to fermionic operators $\hat c^{(\dagger)}_k, \hat c^{(\dagger)}_{\odot,k}$, the corresponding Hamiltonian in the periodic\hyp ladder representation~\cite{Wilke23} is given by $\hat H_\mathrm{wheel}^\mathrm{noise} = \hat \Pi_\odot \hat H_\mathrm{lad}^\mathrm{noise} \hat{\Pi}_\odot$ with
\begin{align}
	\hat H _\mathrm{lad}^\mathrm{noise}
	&=
	-\sum_{k\in\mathcal M} \left[ 2t \cos (k) \hat n_k^\nodagger \left( 1 - \frac{2}{L} \hat{n}_{\odot, k=0} \right) + \Tilde{s} \left( f_k^\nodagger \hat{c}_k^\dag \hat{c}_{\odot, k=0}^\nodagger + \text{h.c.} \right) \right], \label{eq:app:wheel:disorder}
\end{align}
where the sum is over all momenta $k \in \mathcal M$.
Here, we introduced the Fourier transformed, perturbed ring\hyp to\hyp center hoppings $f_k = \delta_{k,k_0} + \frac{1}{\sqrt L}\sum_j ( \delta s_j / \Tilde{s} ) \mathrm e^{\mathrm i (k_0 - k) j}$.
Considering an ensemble of wheels, one could immediately evaluate the average of $f_k$ using the fact that $\sqrt{\mathrm{Var} [f_k]} = \sigma/\Tilde{s}$ and $\mathrm{E}[f_k] = \delta_{k,k_0}$, which implies a quick convergence to the unperturbed system's spectrum in the limit $\Tilde{s} / \sigma \to \infty$.
However, in our setup a wheel is supposed to represent a single logical qubit.
As a consequence, we have to investigate the spectral properties of~\cref{eq:app:wheel:disorder} for individual realizations of the imperfections $\delta s_j$ at finite $\Tilde s / \sigma$.
From the analysis of the unperturbed Bose\hyp Hubbard wheel it is known that the single\hyp particle gap controls the separation of the many\hyp particle spectrum into trivial and~\gls{BEC} states.
Hence, we start our discussion with an analysis of the perturbed Bose\hyp Hubbard wheel in the single\hyp particle subspace.
In that case, exploiting $\hat{n}_k \hat{n}_{\odot, k=0} = 0$, we can solve~\cref{eq:app:wheel:disorder} for the single\hyp particle eigenstates, which can be divided into two sets.
The first set of eigenstates is characterized by momenta $q\in \{ k_n \in \mathcal M | 0<k_n<\pi \} \equiv \mathcal M^\prime$:
\begin{align}
	\hat{\Psi}_q^\dagger \ket{\varnothing}
	=
	d_q \left( f^\ast_{2\pi - q} \hat{c}_k^\dagger - f^\ast_k \hat{c}_{2\pi - q}^\dagger \right) \ket{\varnothing}\; ,
\end{align}
with single\hyp particle eigenvalues $E_q = -2t \cos{q} \in ]-2t, 2t[$ and normalization $d_q$.
For $L$ even (odd), these $|\mathcal M^\prime| = (L-2)/2$ states ($|\mathcal M^\prime| = (L-1)/2$ states) represent superpositions of plane waves occurring on the outer ring exclusively and their energies are constrained by the usual tight\hyp binding dispersion relation.
The second set of eigenstates non\hyp locally couples excitations on the outer ring with excitations on the inner ring and can be parametrized as
\begin{align}
	\hat{\Psi}_n^\dagger \ket{\varnothing} = d_n \left( \hat{c}_{\odot, k=0}^\dagger + \sum_{k \in \mathcal M} b_{n,k} \hat{c}_k^\dagger \right) \ket{\varnothing} \; .
\end{align}
Here, the parameters $b_{n,k} = - \Tilde s f_k / (E_n + 2t\cos k)$ need to be determined from solving a self\hyp consistent equation for the corresponding single\hyp particle eigenvalues
\begin{align}
	E_n = \Tilde{s}^2 \sum_{k \in \mathcal M} \frac{|f_k|^2}{E_n + 2t \cos{k}} \equiv \Tilde{s}^2 g(E_n)\; . \label{eq:wheel:disorder:sp:g}
\end{align}
Even though this equation does not allow for an analytic solution, it is possible to gain further insights by closer analyzing the function $g(E)$.
Its domain consists of the set of disjoint intervals $\mathcal I_n = ]-2t \cos{k_n},-2t \cos{k_{n+1}}[$ and the two marginal intervals $\mathcal I_- = ]-\infty, -2t[$ and $\mathcal I_+ = ]2t, \infty[$.
In each interval, $g(E)$ is continuous and strictly decreasing with poles located at the boundaries, which are given by the tight\hyp binding eigenvalues $E_n = -2t\cos k_n$.
To illustrate the structure of $g(E)$, in~\cref{fig:wheel:disorder:sp:g} we show a realization for a certain choice of perturbations $\delta s_j$ and $L=6$ sites on the outer ring.
The branches of $g(E)$ in the intervals $\mathcal I_n, \mathcal I_\pm$ are shown by the blue lines.
The solutions to~\cref{eq:wheel:disorder:sp:g} are located at the intersections between the function $y(E) = E$, i.e., the main diagonal in~\cref{fig:wheel:disorder:sp:g} plotted as green line, and $\Tilde{s}^2 g(E)$.
They are marked as green dots, shown in the left panel.
Since both functions, $y(E)$ and $g(E)$, are strictly increasing and decreasing in every interval, respectively, there are unique solutions $E_n \in \mathcal I_n$ and $E_\pm \in \mathcal I_\pm$, i.e., for $L$ even (odd), there are $(L+4)/2$ eigenvalues ($(L+3)/2$ eigenvalues) contributed from the second set of Eigenstates, which confirms that the single\hyp particle problem is indeed solved completely by determining the solutions $E_n$ to~\cref{eq:wheel:disorder:sp:g}.
The marginal energies $E_\pm$ are of primary interest because they determine the size of the single\hyp particle gaps.
For that reason, we establish an upper bound for the solution $E_n = \Tilde s^2 g(E_n)$ with $E_n < -2t$ by estimating $g(E_n) < |f_k|^2 / (E_n + 2t \cos{k})$ for all $k \in \mathcal M$.
Similarly, a lower bound can be introduced via $g(E_n) > \sum_{k \in \mathcal M} |f_k|^2 / (E_n + 2t)$, yielding bounds for the lower marginal energy
\begin{align}
	E_- < -t \cos{k} - \sqrt{(t \cos{k})^2 + \Tilde{s}^2 |f_k|^2}\;,
	\label{eq:wheel:disorder:sp:e_minus_upper}
\end{align}
for all $k \in \mathcal M$ and
\begin{align}
	E_- > -|t| - \sqrt{t^2 + \Tilde{s}^2 \sum_{k \in \mathcal M} |f_k|^2} \;.
	\label{eq:wheel:disorder:sp:e_minus_lower}
\end{align}
The bounds on $\Tilde{s}^2 g(E_-)$ and $E_-$ are shown in the right panel of~\cref{fig:wheel:disorder:sp:g} where we evaluated the upper bound at $k=k_0$, yielding the best approximation.
Accordingly, the marginal single\hyp particle energy $E_+$ can be bounded and we show the resulting best bounds in the right panel of~\cref{fig:wheel:disorder:sp:g}.
An important result of these estimations is that the marginal energies $E_\pm$ are gapped out from the bulk of the single\hyp particle spectrum $\propto \lvert\Tilde{s} f_{k_0}\rvert$.
\begin{figure}[t]
	\centering
	\ifthenelse{\boolean{buildtikzpics}}%
	{%
		\input{figures/single_particle_stats_noise}%
	}%
	{%
		\includegraphics{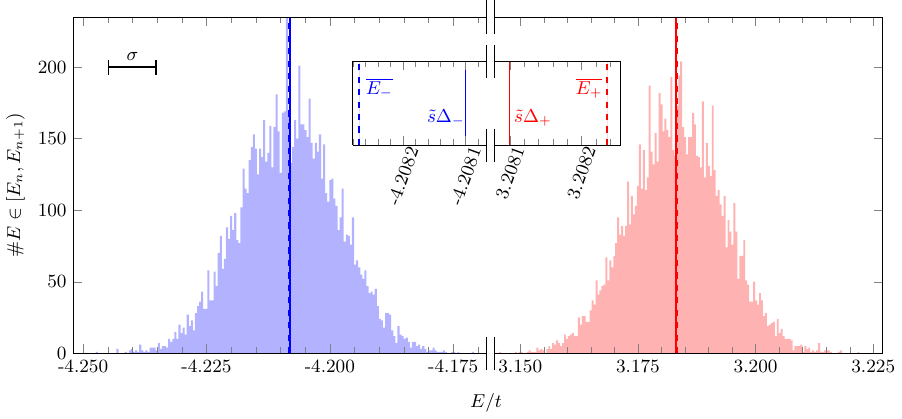}%
	}%
	\caption{
		Distributions of the marginal energies $E_\pm$ for $10^4$ independent realizations $\left\{ \delta s_j \right\}$ of the Bose\hyp Hubbard wheel with $L=6$ lattice sites on the outer ring, $k_0 = \pi/3$, $s/t = 1.5$ and random imperfections $\delta s_j$ with $\sigma/t = 0.01$.
		The sample\hyp means for the single\hyp particle eigenvalues $E_+$ ($E_-$) are represented by the red (blue) dashed lines.
		The solid lines indicate the marginal single\hyp particle eigenvalues of the unperturbed system separating from the bulk spectrum $\propto \Tilde{s}$.
		The means of the perturbed and the unperturbed marginal energies are nearly on top of each other (we show an enlarged excerpt in the insets), even for the rather moderate choices of the ring\hyp to\hyp center hopping and the width of the distribution of the imperfections.
	}
	\label{fig:sp_statistics}
\end{figure}
\newline
The established bounds on the marginal single\hyp particle energies explicitely depend on the realization of the perturbed ring\hyp to\hyp center hoppings $s+\delta s_j$ via $f_k$.
Thus, controlling the width $\sigma$ of the distribution of the imperfections $\delta s_j$ is important.
A justification for assuming $\sigma/s \ll 1$ is provided in \ref{app:noisy-couplings}.
We statistically analyze $E_\pm$ neglecting quadratic and higher orders in $\delta s_j$ on the right-hand side of~\cref{eq:wheel:disorder:sp:g}.
Thereby, we obtain two solutions $\Tilde{E}_\pm$, which fulfill the self-consistency condition up to quadratic order in $\sigma$, yielding expectation values given in \cref{eq:wheel:disorder:Etilde}, which correspond to the marginal single\hyp particle energies of the isotropic Bose\hyp Hubbard wheel.
Furthermore, the variance can be computed in this approximation, yielding $\mathrm{Var}[\Tilde{E}_\pm] \in \mathcal{O}(\sigma^2)$.
Evaluating the boundaries' variances from~\cref{eq:wheel:disorder:sp:e_minus_upper,eq:wheel:disorder:sp:e_minus_lower} directly, yields the same scaling in $\sigma$, which justifies the simplification of neglecting higher orders in the imperfections $\delta s_j$.
As a consequence, the solutions $\Tilde{E}_\pm$ approximate the marginal energies $E_\pm$ for sufficiently large ratios $\Tilde{s} / t$ and both approach the marginal single\hyp particle eigenvalues of the unperturbed Bose\hyp Hubbard wheel quickly.
This is demonstrated in~\cref{fig:sp_statistics} by incorporating the values of $E_\pm$ from $10^4$ independently implemented systems.
\newline
The analysis of the stability of the single\hyp particle spectrum against perturbations of the wheel\hyp to\hyp center hopping suggests that the results for the many\hyp body problem of the unperturbed wheel can be adopted to the perturbed Bose\hyp Hubbard wheel, too.
Here, we employ the Fourier transformed, perturbed wheel-to-center hoppings $f_k$ again and separate the contribution from the noise, which allows to rewrite the projected ladder representation of the perturbed Bose\hyp Hubbard wheel as
\begin{align}
	\hat{\Pi}_\odot^\nodagger \hat{H}_\mathrm{lad}^\mathrm{noise} \hat{\Pi}_\odot^\nodagger &= \hat{\Pi}_\odot^\nodagger \hat{H}_\mathrm{lad} \hat{\Pi}_\odot^\nodagger + \frac{1}{\sqrt{L}} \hat{\Pi}_\odot^\nodagger \hat{V} \hat{\Pi}_\odot^\nodagger, ~ \mathrm{with} ~ \hat{V} \equiv - \sum_{k \in \mathcal M} \left( \sum_{j=0}^{L-1} \delta s_j e^{i(k_0 - k)j} \hat{c}_k^\dagger \hat{c}_{\odot, k=0}^\nodagger + \mathrm{h.c.} \right) \; .
\end{align}
We collected all contributions $\propto \delta s_j$ in $\hat V$, which we treat as a perturbation in order to analyze the stability of the branches in the many\hyp body spectrum of the perturbed Bose\hyp Hubbard wheel.
We find that the expectation values of the many\hyp body eigenvalues $E_\mathrm{noise}$ approach the many\hyp body eigenvalues $E$ of the unperturbed system extremely fast.
Moreover, the variance is in $\mathcal{O}(L^{-3})$ for perturbed many\hyp body eigenstates corresponding to unperturbed eigenstates in the central branch.
For those states that correspond to the perturbed eigenstates in the separating branches, we find that the variances are dominated by the single\hyp particle behaviour and the coefficients $c_{(+/-)}$, fulfilling $c_{(+/-)} \overset{\Tilde{s}/t \to \infty}{\longrightarrow} 1$, do not depend on $\sigma$.
\subsection{\label{app:noisy-couplings}Deriving noisy couplings from experimental constraints}
\begin{figure}%
	\ifthenelse{\boolean{buildtikzpics}}%
	{%
		\input{figures/double_harmonic_potential}
	}
	{
		\includegraphics{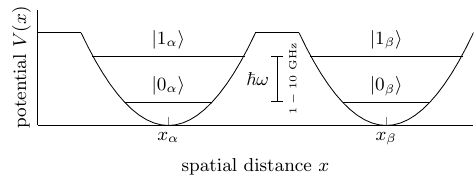}
	}
	\caption%
	{
		Truncated double-harmonic potential for modeling a single \gls{SCQ} coupling to the cavity.
		In the limit of infinitely separated potential wells we approximate the system's eigenstates with the two lowest bound states of isolated harmonic oscillators $\ket{0_i}$ and $\ket{1_i}$ ($i=\alpha,\beta$).
		The hybridization $\braket{1_\alpha,0_\beta|0_\alpha,1_\beta}$ constitutes an approximation to the ring\hyp to\hyp center hopping $s(a)$ as a function of the spatial distance $a = x_\beta - x_\alpha$ of the potential wells.
	}%
	\label{fig:double_harmonic_potential}
\end{figure}
\begin{figure}[t]%
	\centering%
	\ifthenelse{\boolean{buildtikzpics}}%
	{%
		\input{figures/toy_model_spectrum}
	}
	{
		\includegraphics{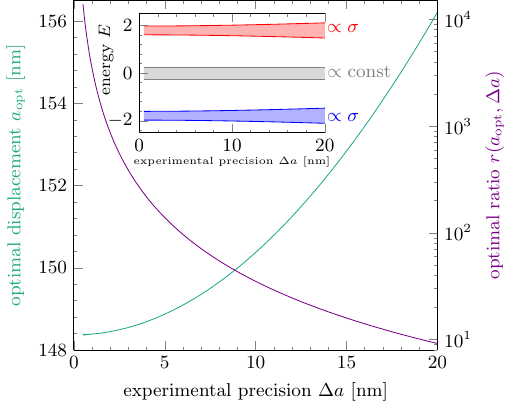}
	}
	\caption{%
		\label{fig:wheel:disorder:spectrum:toymodel}
		Numerically optimized mean distance $a_\mathrm{opt}$ and optimal ratio $r(a_{opt}, \Delta a)$, depending on the experimental precision of the transmon qubit location $\Delta a$.
		The displayed data results from system sizes $L = 20$ and $N=4$ particles, as well as a modulation of $k_0 = \pi / 2$.
		For readability reasons, the value of $s = 0.4$ is kept constant with a typical transmon frequency of $\omega / 2\pi = \unit[5]{GHz}$ and $s = 12$.
		The inset plot shows the broadening of the separated many-particle clusters corresponding to an interval of three standard deviations, $E^\mathrm{noise}_\mathrm{\pm} \pm 3 \sqrt{c_\pm} \sigma$.
	}
\end{figure}

%
While there are many sources for experimental losses, in our case a relevant quantity is the noise introduced by imperfections of the displacement of the~\glspl{SCQ} from the cavity.
In order to estimate the magnitude of the resulting perturbations we consider a single realization of a \gls{SCQ} coupled to a cavity, which can be described by a Jaynes-Cunnings model in the regime where only the two lowest-energy qubit states are relevant~\cite{Blais21}.
Assuming a small qubit\hyp cavity detuning and that the effective qubit\hyp cavity coupling is large, we can neglect states with more than one excitation in the cavity such that for the following estimate we assume a system of two harmonic oscillators truncated to the two lowest eigenstates as depicted in~\cref{fig:double_harmonic_potential}.
For that model we can compute the hybridization between qubit and cavity in terms of the oscillator eigenstates, as a function of the spatial distance $a$.
At resonance where the level\hyp spacing in both oscillators is identical, $\omega_\alpha = \omega_\beta = \omega$, we can then deduce the ring\hyp to\hyp center hopping amplitude from the overlap between the ground state $\ket{0_\alpha}$ of one oscillator and the excited state $\ket{1_\beta}$ of the other oscillator
\begin{align}
	s(a)\approx
	\hbar \omega \braket{1_\alpha,0_\beta | 0_\alpha, 1_\beta} 
	= \hbar \omega \left( \frac{m_e \omega}{2\hbar} a^2 - 1 \right) \mathrm{e}^{-\frac{m_e \omega}{4\hbar} a^2} \; . \label{eq:hopping_toy_model}
\end{align}
We can now introduce random variations $\delta a_j$ describing experimental imperfections of $a$ so that $a \rightarrow a+\delta a_j \equiv a_j$ for each qubit $j$ coupled to the cavity.
We assume the $\delta a_j$ to be normal distributed, independent, random variables, $\delta a_j \sim \mathcal{N}(0, \Delta a^2)$.
Here, $\Delta a$ is fixed by the experimental precision.
Evaluating the first moment of~\cref{eq:hopping_toy_model} with respect to the distribution of the $a_j$ yields an estimation for the ring\hyp to\hyp center hopping $s(a,\Delta a) = \langle\langle s \rangle\rangle_{a,\Delta a}$ while the second moment constitutes the standard deviation of the perturbations $\sigma^2(a,\Delta a) = \langle\langle s^2 \rangle\rangle_{a,\Delta a} - \langle\langle s \rangle\rangle^2_{a,\Delta a}$.
The previous considerations enable us to numerically compute the optimal displacement $a_\mathrm{opt}$ of the \glspl{SCQ} from the cavity.
We determined $a_\mathrm{opt}$ such that the ratio $r(a_\mathrm{opt}, \Delta a) = s(a_\mathrm{opt},\Delta a)/\sigma(a_\mathrm{opt},\Delta a)$ is maximized, assuming precisions for positioning the qubits of $\Delta a \in [0.5\,\mathrm{nm}, 20\mathrm{nm}]$.
In~\cref{fig:wheel:disorder:spectrum:toymodel} $a_\mathrm{opt}$ is shown as a function of the precisions $\Delta a$ by the green curve, which is strictly increasing.
The corresponding optimal ratio, shown by the purple curve, demonstrates that for practically feasible precisions $\Delta a \sim \mathcal O(1\,\mathrm{nm})$ large ratios $r(a_\mathrm{opt}, \Delta a) \sim \mathcal O(10^4)$ are in reach.
We conclude that within the discussed approximations, the effect of model\hyp imperfections in the ring\hyp to\hyp center hopping can be suppressed efficiently, if the \glspl{SCQ} are placed in the optimal distance $a_\mathrm{opt}$ from the cavity.
In fact, for the optimal distance, the broadening of the characteristic clusters in the many\hyp body spectrum of the unperturbed~\gls{BHW}, caused by experimental imperfections, is of the order $s\times 10^{-4}$.

\section{Solution strategy of the Bose\hyp Hubbard wheel with control qubit}
\begin{figure}
	\subfloat[\label{fig:gap:E1E2}]%
	{%
		\ifthenelse{\boolean{buildtikzpics}}%
		{%
			\input{figures/heatmap_E1_E2}%
		}%
		{%
			\includegraphics{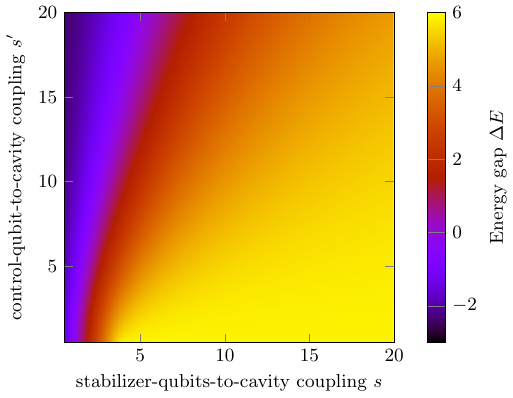}%
		}%
	}%
	\hfill%
	\subfloat[\label{fig:gap:EcE2}]%
	{%
		\ifthenelse{\boolean{buildtikzpics}}%
		{%
			\input{figures/heatmap_Ec_E2}%
		}%
		{%
			\includegraphics{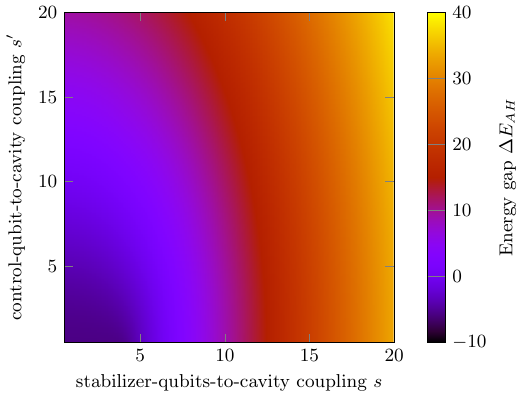}%
		}%
	}%
	\caption
	{
		Energy differences between specific points in the two lowest clusters (i.e., $\nu=0$ and $\nu=1$) of the many\hyp body spectrum for different $s$ and $s^\prime$ for $L = 6, N = 3$. 
		In \subfigref{fig:gap:E1E2} we plot $\Delta E = \min\limits_\mu E_{0\mu} (N+1,N_{k_0}=2) - \max\limits_\mu E_{0\mu} (N, N_{k_0}=1)$, i.e., the separation between the lower edge of the $N_{k_0}=2$ and the upper edge of the $N_{k_0}=1$ cluster.
		This energy differences characterizes the gap in the effective, logical two\hyp level system.
		\subfigref{fig:gap:EcE2} shows $\Delta E_{AH} = \min\left(\min\limits_{\mu,N_{k_0}=0,3} E_{0\mu} (N+1,N_{k_0}), \min\limits_{\mu,N_{k_0}=1,2} E_{1\mu} (N+1,N_{k_0})\right) - \max_{\mu,N_{k_0}=1,2} E_{0\mu} (N, N_{k_0})$, i.e., the separation between the lower edge of the central many\hyp body eigenstate cluster and the upper edge of the lower many\hyp body eigenstate cluster.
		This energy differences characterizes the separation of the effective, logical two\hyp level system from the remaining many\hyp body eigenstates, which can be compared to the anharmonicity of the constituting~\gls{SCQ}.
	}
\end{figure}
Let us now derive the matrix representation of the extended system, in which a newly introduced control qubit, denoted by $c$, couples to the center site of the wheel with amplitude $s^\prime$.
The Hamiltonian of interest is given by
\begin{align}%
	\hat H_{\mathrm{qubit}} = \hat H_{\mathrm{wheel}} + \hat H_c = -t \sum_{j=0}^{L-1} \left( \hat h^\dagger_j \hat h^\nodagger_{j+1} + \mathrm{h.c.} \right) -%
	\sum_{j=0}^{L-1} \left( s_j\hat h^\dagger_j \hat h^\nodagger_\odot + \mathrm{h.c.} \right)  + s^\prime (\hat h^\dagger_\odot \hat h^\nodagger_c +  \mathrm{h.c.}) + \mu_c \hat n_c\;.
	\label{eq:sm:hubb-wheel-probe}%
\end{align}%
Here, $\hat h^{(\dagger)}_c$ denotes the annihilation (creation) of a \gls{HCB} on the control qubit.
We introduce $\hat N_{k_0} = \hat n_c + \hat n_{k_0}$, the combined occupation of the $k_0$ mode, introduced in the section above, and the occupation of the control qubit.
Since $\left[ \hat H, \hat N_{k_0}\right] = 0 $, the Hamiltonian can be block\hyp diagonalized in $N_{k_0}$, $\braket{\mathrm{FS}_N|\hat H_\mathrm{qubit} | \mathrm{FS}_N} = \bigoplus_{N_{k_0}} \hat h(\mathrm{FS}_N, N_{k_0})$.
After re-sorting the blocks and grouping together those with the same total number of occupied modes in the Slater determinants and the $N_{k_0}$ sector, we obtain a $8\times 8$ matrix for a given $N$ made up of blocks for each $N_{k_0}$ with $N-N_{k_0}$ occupied modes in the Slater determinants, see \cref{fig:sm:subspaces} (right).
Hence, we work in the basis
\begin{align}%
	\ket{n_c}\ket{n_{k_0}}\ket{FS_{N-N_{k_0}}}\;,
	\label{eq:sm:probe_basis}%
\end{align}%
with possible values $n_c = 0,1$ and  $n_{k_0} = 0,\pm$ (denoting the $n_{k_0}=1$ eigenstates), $2$, where the remaining eigenvalue problem of~\cref{eq:sm:hubb-wheel-probe} can be solved in each block, individually.
The matrix elements of each block are obtained in the following by letting \cref{eq:sm:hubb-wheel-probe} act on the states \cref{eq:sm:probe_basis}.
For the $N_{k_0} = 0,3$ sectors, the matrix elements are given by $E_{0}(FS_{N})$ and  $E_{0}( FS_{N-3})$ respectively.
For the $N_{k_0} = 1$ sectors we have
\begin{align}
	&\hat H_c \ket{n_c = 0} \ket{\pm} \ket{FS_{N-1}} \notag\\ & = s^\prime\left( \hat h^\dagger_c \hat h^\nodagger_\odot + \hat h^\dagger_\odot \hat h^\nodagger_c\right) \ket{n_c = 0} \left(d^\pm_0\ket{n_{k_0,O} = 1}\ket{n_\odot = 0} + d^\pm_1\ket{n_{k_0,O} = 0}\ket{n_\odot = 1} \right)\ket{FS_{N-1}}\notag\\
	& = s^\prime d^\pm_1 \ket{n_c = 1}\ket{n_{k_0,O} = 0}\ket{n_\odot = 0}\ket{FS_{N-1}} = s^\prime d^\pm_1\ket{n_c = 1}\ket{0}\ket{FS_{N-1}}\notag \;.
\end{align}
It follows that 
\begin{align}
	\bra{FS_{N-1}}\bra{0}\bra{n_c = 1}\hat H_c \ket{n_c = 0} \ket{\pm} \ket{FS_{N-1}} = s^\prime d^\pm_1\;.
\end{align}
All other matrix elements can be computed in a similar manner.
Note that $\hat H_{\mathrm{wheel}}$ only has diagonal terms.
Due to the orthogonality of the projected wheel eigenstates all other matrix elements vanish.
$\hat H_c$ does not contribute diagonal terms as it changes the particle number on the control qubit and hence only connects states with different $n_c$.
The entire $N_{k_0} =1$ subspace in matrix representation is given by 
\begin{equation}
	\hat h(\mathrm{FS}_{N-1},N_{k_0} = 1) = 
\begin{pmatrix}
	E_{0}( \mathbf{k}_{N-1}) + \mu_c& s^\prime (d^+_1) & s^\prime (d^-_1)\\
	s^\prime (d^+_1)^* & E_{1_+}( \mathbf{k}_{N-1})& 	0\\
	s^\prime (d^-_1)^*& 0 &  E_{1_-}( \mathbf{k}_{N-1})
	\label{eq:sm:Nk01_mat}
\end{pmatrix}\;.
\end{equation}
For the $N_{k_0} = 2$ sectors we have
\begin{align}
	&\hat H_c \ket{n_c = 1} \ket{\pm} \ket{FS_{N-2}}\notag\\
	& = s^\prime\left( \hat h^\dagger_c \hat h^\nodagger_\odot + \hat h^\dagger_\odot \hat h^\nodagger_c\right) \ket{n_c = 1} \left(d^\pm_0\ket{n_{k_0,O} = 1}\ket{n_\odot = 0} + d^\pm_1\ket{n_{k_0,O} = 0}\ket{n_\odot = 1} \right)\ket{FS_{N-2}}\notag\\
	& = s^\prime d^\pm_0 \ket{n_c = 0}\ket{n_{k_0,O} = 1}\ket{n_\odot = 1}\ket{FS_{N-2}} = s^\prime d^\pm_0\ket{n_c = 0}\ket{2}\ket{FS_{N-2}} \notag\;,
\end{align}
from which follows
\begin{align}
	\bra{FS_{N-2}}\bra{2}\bra{n_c = 0}\hat H_c \ket{n_c = 1} \ket{\pm} \ket{FS_{N-2}} = s^\prime d^\pm_0\;.
\end{align}
The entire $N_{k_0} =2 $ subspace in matrix representation is given by 
\begin{equation}
	\hat h(\mathrm{FS}_{N-2},N_{k_0} = 2) = 
	\begin{pmatrix}
		E_{1_-}( \mathbf{k}_{N-2}) + \mu_c& 0 & s^\prime (d^-_0)^*\\
		0 & E_{1_+}( \mathbf{k}_{N-2}) + \mu_c& 	s^\prime (d^+_0)^*\\
		s^\prime d^-_0 & s^\prime d^+_0 & 	E_{2}( \mathbf{k}_{N-2})
		\label{eq:sm:Nk02_mat}
	\end{pmatrix} \;.
\end{equation}
These results are then used for numerical diagonalization.
Arbitrary states from the respective sectors $N_{k_0}$ in a system of $N$ particles are denoted as follows
\begin{align}
	&\ket{E_{\mu} (N,N_{k_0} = 0) }= \ket{\varnothing}\ket{FS_N}_\mu\\
	&\ket{E_{\nu\mu} (N,N_{k_0} = 1)} = \left[ v^{\nu\mu}_0\ket{1,0} +v^{\nu\mu}_+\ket{0,+} + v^{\nu\mu}_-\ket{0,-}\right] \ket{FS_{N-1}}_\mu\\
	&\ket{E_{\nu\mu} (N,N_{k_0} = 2)} = \left[ w^{\nu\mu}_-\ket{1,-} +w^{\nu\mu}_+\ket{1,+} + w^{\nu\mu}_2\ket{0,2}\right] \ket{FS_{N-2}}_\mu\\
	&\ket{E_{\mu} (N,N_{k_0} = 3)} = \ket{1,2}\ket{FS_{N-3}}_\mu
	\label{eq:sm:wheel-probe-eigenstates}
\end{align}
where $\mu\in\binom{L-1}{N-N_{k_0}}$ enumerates the Slater determinant outside of the $N_{k_0}$ subspace (for simplicity, this index has been dropped in the main text and the states are written as $	\ket{E_\nu (N, N_{k_0}) } = \ket{\nu_{N_{k_0}}} \otimes \ket{\mathrm{FS}_{N-N_{k_0}}}$ with $\nu_{N_{k_0}} = 0,\dots, d_{N_{k_0}}-1$ where $d_{N_{k_0}} $ denotes the dimension of each $N_{k_0}$ block) and we use the abbreviation $\ket{n_c,n_{k_0}}$.
For the non\hyp trivial $N_{k_0} = 1,2$ sectors, there is an additional index $\nu$ denoting all three eigenstates for a given $\mu$.
Note that each of the corresponding energies can be directly identified with one of the three branches in the spectrum and $\nu = 0$ corresponds to the low\hyp lying BEC sector which is of particular interest.
Given the many\hyp body energies, we can characterize the typical energy scales relevant for the logical qubit setup.
First, there is the energy gap between the two logical qubit states
\begin{equation}
	\Delta E = \min\limits_\mu E_{0\mu} (N+1,N_{k_0}=2) - \max\limits_\mu E_{0\mu} (N, N_{k_0}=1) \; ,
\end{equation}
which we show exemplary in~\cref{fig:gap:E1E2} for a fixed system size and particle number, varying the couplings $s, s^\prime$.
The second relevant energy scale is the separation of the two logical qubit states from the remaining part of the many\hyp body spectrum
\begin{equation}
	\Delta E_{AH} = \min\left(\min\limits_{\mu,N_{k_0}=0,3} E_{0\mu} (N+1,N_{k_0}), \min\limits_{\mu,N_{k_0}=1,2} E_{1\mu} (N+1,N_{k_0})\right) - \max_{\mu,N_{k_0}=1,2} E_{0\mu} (N, N_{k_0}) \; .
\end{equation}
This quantity is shown in~\cref{fig:gap:EcE2} using the same parameters as before.
The structure of the eigenstates of the reduced Hamiltonians $\hat h(\mathrm{FS}_{N-N_{k_0}},N_{k_0})$, as well as the occupation of the control qubit can also be understood by considering the control qubit as perturbation in $s^\prime / \Tilde s$.
Expanding $\hat h(\mathrm{FS}_{N-N_{k_0}},N_{k_0})$ in the tensor product basis of eigenstates of the reduced wheel $\ket{E_{n_{k_0}}(\mathrm{FS}_{N-n_{k_0}})}$ and the control qubit $\ket{n_c}$, one readily finds that the first non\hyp trivial correction appears in second order $s^\prime / \Tilde s$.
In the following we will temporarily drop the Slater determinants $\mu$.
The correction to the odd parity $N_{k_0}=1$ eigenstate is given by
\begin{align}
		\ket{E_{1} (N,N_{k_0} = 1)}
		&= \ket{1,0} + \mathcal O\left(\frac{s^\prime}{\Tilde s}\right) \sum_{n_{k_0}=\pm}\ket{0,n_{k_0}} \; .
\end{align}
For the even parity eigenstates with $\nu = 0,2$ one equivalently finds
\begin{align}
	\ket{E_{\nu} (N,N_{k_0} = 1)}
	&= \ket{0,\pm} + \mathcal O\left(\frac{s^\prime}{\Tilde s}\right) \ket{0,0} \; .
\end{align}
Therefore, the probe\hyp site occupations in the eigenstates of the $N_{k_0} = 1$ sector exhibit second\hyp order corrections in $s^\prime / \Tilde s$:
\begin{align}
		\bra{E_{1} (N,N_{k_0} = 1)}\hat n_c	\ket{E_{1} (N,N_{k_0} = 1)} &= 1 - \mathcal O \left(\left(\frac{s^\prime}{\Tilde s}\right)^2 \right) \;,\\
		\bra{E_{\nu} (N,N_{k_0} = 1)}\hat n_c	\ket{E_{\nu} (N,N_{k_0} = 1)} & = \mathcal O \left(\left(\frac{s^\prime}{\Tilde s}\right)^2 \right),
\end{align}
Similarly, for the $N_{k_0}=2$ sector the probe\hyp site occupations evaluate to
\begin{align}
	\bra{E_{1} (N,N_{k_0} = 2)}\hat n_c	\ket{E_{1} (N,N_{k_0} = 2)} &= \mathcal O \left(\left(\frac{s^\prime}{\Tilde s}\right)^2 \right) \;,\\
	\bra{E_{\nu} (N,N_{k_0} = 2)}\hat n_c	\ket{E_{\nu} (N,N_{k_0} = 2)} & = 1- \mathcal O \left(\left(\frac{s^\prime}{\Tilde s}\right)^2 \right)\;,
\end{align}
\section{Fidelity of $X$\hyp gate application}
In the following, we will consider systems at half filling $N = L/2$ and make use of $\hat h^\dagger_c = \hat c^\dagger_c - \mathcal O \left(\left(\frac{s^\prime}{\Tilde s}\right)^2\right)$.
For later convenience and using \cref{eq:sm:wheel-probe-eigenstates} we compute
\begin{align}
	&\hat c^\dagger_c \ket{E_\mu(N,N_{k_0}=0)} = \ket{n_c = 1}\ket{FS_N}_\mu\notag\\
	&\hat c^\dagger_c \ket{E_{\nu\mu}N,N_{k_0}=1}= \left[ v^{\nu\mu}_+ \ket{n_c = 1, +} + v^{\nu\mu}_- \ket{n_c = 1, -}\right] \ket{FS_{N-1}}_\mu\notag \\
	&\hat c^\dagger_c \ket{E_{\nu\mu}(N,N_{k_0}=2)} = w^{\nu\mu}_2 \ket{n_c = 1, 2}\ket{FS_{N-2}}_\mu \notag\\
	&\hat c^\dagger_c \ket{E_{\mu}(N,N_{k_0}=3)} = 0 \notag
\end{align}

We define the $X$\hyp gate fidelity at zero temperature as the probability to create a state in the lower branch of the $N_{k_0} = 2$ sector of a system realization with $N+1$ particles by exciting a state in the lower branch of the $N_{k_0} = 1$ sector of a system with $N$ particles,
\begin{align}
	F \equiv \abs{\bra{E_{0\mu}(N+1,N_{k_0}=2)}\hat c^\dagger_c\ket{E_{0\mu}(N,N_{k_0}=1)}}^2 =
	 \abs{(\tilde w^{0\mu}_+)^*v^{0\mu}_+ + (\tilde w^{0\mu}_-)^*v^{0\mu}_-}^2\;.
	\label{eq:fidelity}
\end{align}
The tilde highlights the fact that we are looking at two different initializations of the wheel\hyp probe system with $N $ and $(N+1)$  particles denoted by coefficients without and with tilde.
For the finite temperature treatment we introduce the thermal density operator of the system
\begin{align}
	\hat \rho (T,N) = 
	\frac{1}{Z_\beta} \sum_i e^{-\beta E_i^N}\ket{\Psi_i(N)}\bra{\Psi_i(N)} 
	= \frac{1}{Z_\beta} \sum_{N_{k_0}=0}^3\sum_\mu^{\binom{L-1}{N-N_{k_0}}}\sum_\nu^{d_{N_{k_0}}} e^{-\beta E_{\nu,\mu}^N(N_{k_0})} \ket{E_{\nu\mu}(N,N_{k_0})}\bra{E_{\nu\mu}(N,N_{k_0})}
	\label{eq:sm:thermal_density}
\end{align}
with $\beta = 1/k_BT$, $d(N_{k_0})$ denotes the dimension of the $N_{k_0}$ subspace (we will drop the index $\nu$ in the case $d(N_{k_0}) = 1$) and the partition function 
\begin{align}
Z_\beta = \sum_{N_{k_0}=0}^3\sum_\mu^{\binom{L-1}{N-N_{k_0}}}\sum_\nu^{d_{N_{k_0}}} e^{-\beta E_{\nu\mu}^N(N_{k_0})}\;.
\end{align}
The energies are normalized to the ground state energy.
To examine the effect of an excitation on the control qubit we consider the density operator
\begin{align}
	&\hat \rho_1(T) = \hat c^\dagger_c \hat \rho (T) \hat c^\nodagger_c\notag
	 = \frac{1}{Z_\beta}
	 \sum_\mu^{\binom{L-1}{N}} e^{-\beta E^N_{\mu}(0)}  \ket{n_c= 1}\ket{FS_N}_\mu\bra{n_c=1}\bra{FS_N}_\mu \\
	& + \sum_\mu^{\binom{L-1}{N-1}}\sum_{\nu}^{d_1} e^{-\beta E^N_{\nu\mu}(1)} \notag
	 \left[v^{\nu\mu}_+ \ket{n_c = 1,+} + v^{\nu\mu}_- \ket{n_c = 1,-}\right] \ket{FS_{N-1}}_\mu
	 \left[(v^{\nu\mu}_+)^* \bra{n_c = 1,+} + (v^{\nu\mu}_-)^* \bra{n_c = 1,-}\right] \bra{FS_{N-1}}_\mu\\
	&+ \sum_\mu^{\binom{L-1}{N-2}}\sum_{\nu}^{d_2} e^{-\beta E^N_{\nu\mu}(2)} \abs{w^{\nu\mu}_2}^2 
	 \ket{n_c = 1,2}\ket{FS_{N-2}}_\mu \bra{n_c = 1,2}\bra{FS_{N-2}}_\mu ]\;.
	\label{eq:sm:exitation}
\end{align}
The fidelity $F(T)$ is then computed as the probability to find any state $\ket{E_{0\mu}(N+1,N_{k_0}=2)}$ in the low\hyp lying \gls{BEC} sector of the wheel\hyp control qubit system filled with $N+1$ particles after the excitation. 
After some algebra using the orthogonality of Slater determinants we arrive at
\begin{align}
	F(T) &\equiv \sum_\mu^{\binom{L-1}{N-1}} \bra{E_{0\mu}(N+1, N_{k_0} = 2)}\hat c^\dagger_c \hat \rho (T) \hat c^\nodagger_c \ket{E_{0\mu} (N+1, N_{k_0}=2)}\notag\\
	&=\frac{1}{Z} \sum_\mu^{\binom{L-1}{N-1}} \sum^{d_1}_{\nu} e^{-\beta E^N_{\nu\mu}(1)} \abs{ \left[ (\tilde w^{0\mu}_-)^*\bra{n_c=1,-} +(\tilde w^{0\mu}_+)^*\bra{n_c=1,+} + (w^{0\mu}_2)^*\bra{n_c =0,2}\right]\left[v^{\nu\mu}_+\ket{n_c=1,+} + v^{\nu\mu}_-\ket{n_c=1,-}\right]}^2\notag \\
	&= \frac{1}{Z} \sum_\mu^{\binom{L-1}{N-1}} \sum^{d_1}_{\nu} e^{-\beta E^N_{\nu\mu}(1)} \abs{(\tilde w^{0\mu}_+)^* v^{\nu\mu}_+ + (\tilde w^{0\mu}_-)^* v^{\nu\mu}_-}^2\;.
	\label{eq:fideliT}
\end{align}
Only the contributions from the $N_{k_0}=1$ sector are relevant since the Slater determinants have to coincide with those of  $\ket{E_{0\mu} (N+1,N_{k_0}=2)}$, which cannot happen in any other sector, because they have a different number of modes in their Slater determinants.
\section{Sampling Scheme}
The system is prepared in a state described by \cref{eq:sm:thermal_density} and a measurement of the control qubit occupation is performed after exciting the control qubit, i.e. we compute the probability to find the system in the states $\ket{n_c=1}\ket{FS_N}_\mu, \ket{n_c=1,+}\ket{FS_{N-1}}_\mu, \ket{n_c=1,-}\ket{FS_{N-1}}_\mu, \ket{n_c=1,2}\ket{FS_{N-2}}_\mu$ after the excitation.
\begin{align}
	&\bra{n_c = 1}\hat\rho_1(T)\ket{n_c=1} \notag\\
	&= \sum_\mu \bra{n_c=1}\bra{FS_N}_\mu \hat\rho_1(T)\ket{n_c=1}\ket{FS_N}_\mu
	+\sum_\mu \bra{n_c=1,+}\bra{FS_{N-1}}_\mu \hat\rho_1(T)\ket{n_c=1,+}\ket{FS_{N-1}}_\mu + \notag\\
	& \sum_\mu \bra{n_c=1,-}\bra{FS_{N-1}}_\mu \hat\rho_1(T)\ket{n_c=1,-}\ket{FS_{N-1}}_\mu
	+\sum_\mu \bra{n_c=1,2}\bra{FS_{N-2}}_\mu \hat\rho_1(T) \ket{n_c=1,2}\ket{FS_{N-2}}_\mu \notag\\
	& = \frac{1}{Z_\beta} \left[ \sum^{\binom{L-1}{N}}_\mu	e^{-\beta E^N_\mu(0)} +
	 \sum^{\binom{L-1}{N-1}}_\mu	\sum_\nu^{d_1} e^{-\beta E^N_{\nu\mu}(1)}  \left[	\abs{v_+^{\nu\mu}}^2 + \abs{v_-^{\nu\mu}}^2	\right]  +
	  \sum^{\binom{L-1}{N-2}}_\mu	\sum_\nu^{d_2} e^{-\beta E^N_{\nu\mu}(2)} \abs{w_2^{\nu\mu}}^2\right] 
\end{align}
\section{Decoupled qubit manifold in the nonlocal protocol\label{app:nonlocal-protocol}}
	Following the notation from~\cref{eq:sm:pm_coeffs} in~\cref{app:solution_BHW}, the $2\times 2$ blocks are given by
	\begin{align}
		\mathbf H^\mathrm{eff}_\pm
		&=
		\pm\Delta[\mathrm{FS}_N]\mathbb 1_{2\times 2}
		\pm
		\tilde s^\prime
		\begin{pmatrix}
			0 & \bar{d}_1^\pm d_0^\pm\\
			d_1^\pm \bar{d}_0^\pm & 0
		\end{pmatrix} \;, \label{eq:wheel+probe:hamiltonian:non-local:diagonal-blocks} \\
		\mathbf V
		&=
		\tilde s^\prime
		\begin{pmatrix}
			0 & \bar{d}_1^+ d_0^- \\
			d_1^- \bar{d}_0^+ & 0
		\end{pmatrix} \;. \label{eq:wheel+probe:hamiltonian:non-local:offdiagonal-blocks} \\
	\end{align}
	where, in the limit $\tilde s^\prime \ll \tilde s$ we emphasize the opening of a many\hyp body gap 
	\begin{equation}
		2\Delta[\mathrm{FS}_N]
		=
		2 \sqrt{(E[\mathrm{FS}_N]/L + 1)^2 + \tilde s^2} \;,
	\end{equation}
	which is generated by the diverging marginal clusters of many\hyp body eigenstates of the~\gls{BHW}.
	Note that the upper left ($\mathbf H^\mathrm{eff}_+$) and lower right ($\mathbf H^\mathrm{eff}_-$) block act within the $\ket{\pm}$ subspaces, only, whereas the hybridization between the two bright states ($\ket{+}$ and $\ket{-}$) is introduced by $\mathbf V$.
	The resulting hybridization is controlled by an energy scale $\sim \tilde s^\prime$.
	In the limit $\tilde s^\prime \ll \tilde s$, it is possible to decouple the two bright subspaces to second order in $\mathbf V$ via a Schrieffer\hyp Wolff transformation.
	For that purpose, let us denote by $\vec x^\alpha_\pm$ (where $\alpha = 0,1$) the eigenvectors of $\mathbf H^\mathrm{eff}_\pm$:
	\begin{equation}
		\mathbf H^\mathrm{eff}_\pm \vec x^\alpha_\pm = E^\alpha_\pm \vec v^\alpha_\pm \; ,
	\end{equation}
	and define the unitary transformations $\mathbf X_\pm = [\vec x^0_\pm, \vec x^1_\pm]$ as well as $\mathbf X = \mathbf X_+ \oplus \mathbf X_-$.
	Then, there is an anti\hyp hermitian matrix $\mathbf S\in \mathbb V^{4\times 4}_\mathbb{C}$ such that
	\begin{equation}
		\mathrm e^{\mathbf S} \mathbf X^\dagger (H^\mathrm{eff}_Q - \mathcal E[\underline k_N] \mathbb 1_{4\times 4})\mathbf X \mathrm e^{-\mathbf S}
		=
		\mathbf H^Q_+ \oplus \mathbf H^Q_- + \mathcal O(\mathbf V^3) \; ,
	\end{equation}
	and the eigenvalue problem decomposes into a low\hyp energy and a high\hyp energy subspace, in which the $2\times 2$ matrices $\mathbf H^Q_\pm$ constitute the corresponding effective Hamiltonians.
	It is particularly illuminating to investigate the structure of the effective Hamiltonians.
	Let us define the transformed hybridization $\mathbf W = \mathbf X^\dagger_+ \mathbf V \mathbf X_-$, as well as $\mathbf \Lambda_\pm = \operatorname{diag}(E^0_\pm, E^1_\pm)$.
	Then
	\begin{align}
		\mathbf H^Q_+ &= \mathbf \Lambda_+ + \frac{1}{2}\left(\mathbf W \mathbf S^\dagger + \mathbf S \mathbf W^\dagger\right)\;,\\
		\mathbf H^Q_- &= \mathbf \Lambda_- - \frac{1}{2}\left(\mathbf W^\dagger \mathbf S + \mathbf S^\dagger \mathbf W\right)\;,
	\end{align}
	where $S_{nm} = \frac{W_{nm}}{E^n_+ - E^m_-}$ is the generator of the Schrieffer-Wolf transformation.
	Since $\lVert \mathbf V \rVert\sim \mathcal O(\tilde s^\prime)$ the energy corrections are $\sim\mathcal O\left(\frac{(\tilde s^\prime)^2}{\Delta[\underline k_N]}\right)$.
	From $\lvert E^\alpha_\pm \rvert \sim \mathcal O(\Delta[\underline k_N])$ it follows that in the regime $\tilde s^\prime \ll \Delta[\underline k_N]$ the low\hyp lying Hamiltonian forms an effective two\hyp level system
	\begin{equation}
		\mathbf H^Q_-
		=
		\begin{pmatrix}
			\varepsilon^0_- & t_- \\
			\overbar{t}_- & \varepsilon^1_-
		\end{pmatrix}
		\equiv
		\overbar{\varepsilon}_-\mathbb 1_{2\times 2} + 
		\begin{pmatrix}
			\delta & t_- \\
			\overbar{t}_- & -\delta
		\end{pmatrix}
	\end{equation}
	which is energetically protected by a gap $\sim \mathcal O(\Delta[\underline k_N])$, and we introduced the mean energy $\overbar{\varepsilon}_- = (\varepsilon^0_- + \varepsilon^1_-) / 2$ as well as the detuning $\delta = (\varepsilon^0_- - \varepsilon^1_-) / 2$.
	The matrix elements of the effective qubit Hamiltonian are controlled by the parameters $d^\pm_{0/1}$, which are simply the amplitudes of the two bright states of the wheel Hamiltonian $\hat H_\mathrm{wheel}$.
	Consequently, the relevant energy scale $\tilde s/t$ of $\hat H_\mathrm{wheel}$ constitutes a tuning knob, for instance to control the effective hopping and, thereby, provides the possibility to implement coherent logical qubit operations.
	A particular interesting case is $t\equiv 0$, which provides
	\begin{align}
		d^+_0 &= \frac{1}{\sqrt 2}\;, \quad d^-_0 = \frac{1}{\sqrt 2}\;, \\
		d^+_1 &= \frac{1}{\sqrt 2}\;, \quad d^-_1 = -\frac{1}{\sqrt 2}\;.
	\end{align}
	It immediately follows from~\cref{eq:wheel+probe:hamiltonian:non-local:diagonal-blocks} that $E^\pm_0 = E^\pm_1$ as well as $\mathbf W = -\frac{\mathrm i\tilde s^\prime}{2}\mathbf \sigma_y$ and the large\hyp gap limit yields
	\begin{align}
		\mathbf H^Q_-
		&=
		\overbar{\varepsilon}_- \mathbb 1_{2\times 2} - \frac{\mathbf W^\dagger \mathbf W}{2\Delta[\underbar{k}_N]} + \mathcal O\left(\left(\nicefrac{\tilde s^\prime}{\Delta[\underbar k_N]}\right)^2\right) \notag \\
		&=
		\left(\overbar{\varepsilon}_- - \frac{(\tilde s^\prime)^2}{8\Delta[\underline k_N]}\right)\mathbb 1_{2\times 2} + \mathcal O\left(\left(\nicefrac{\tilde s^\prime}{\Delta[\underbar k_N]}\right)^2\right) \;.
	\end{align}
	Remarkably, this establishes $t_- = 0$ with corrections that vanish quadratically in $\tilde s^\prime/\Delta[\underbar k_N]$.
	As a direct consequence, operating the~\gls{BHW} in the $t=0$ regime generates a logical qubit Hamiltonian that is protected against phase errors, because $\delta = 0$, and against bitflip errors because $t_-=0$.
%
\end{document}